%Paper: hep-th/9507140
%From: labastida@vxcern.cern.ch
%Date: Wed, 26 Jul 1995 14:46:28 +0200
%Date (revised): Mon, 16 Oct 1995 10:33:45 +0100
%Date (revised): Thu, 23 Nov 1995 17:32:07 +0100

\input phyzzx.tex
%\draft

 \def\ex{{\hbox{\rm e}}}

  \def\tr{{\hbox{\rm Tr}}}

 \def\vev{vacuum expectation value}

\tolerance=500000 \overfullrule=0pt
\def\np{Nucl. Phys.}
\def\pl{Phys. Lett.}  
\def\pr{Phys. Rev.}  \def\cmp{Comm. Math. Phys.}
   \def\bams{Bull. AMS}   \def\topo{Topology} \def\ijm{Int. J. Math.}  \def\jmp{J. Math. Phys.}  \def\jdg{J. Diff. Geom.}
\def\plms{Proc. London Math. Soc.}
\def\mrl{Math. Res. Lett.}

\def\ex{{\hbox{\rm e}}}     \def\tr{{\hbox{\rm Tr}}}

\def\too{\longrightarrow}

\tolerance=500000 \overfullrule=0pt

\Pubnum={US-FT/21-95 \cr hep-th/9507140}
%\pubnum={US-FT/21-95}
\date={October, 1995 \cr Revised version}
\pubtype={}
\titlepage \title{POLYNOMIAL INVARIANTS FOR $SU(2)$ MONOPOLES}
\author{J.M.F. Labastida\foot{e-mail: labastida@gaes.usc.es} and  M.
Mari\~no\foot{e-mail: marinho@gaes.usc.es}}
\address{Departamento de F\'\i sica
de Part\'\i culas\break Universidade de Santiago\break E-15706 Santiago de
Compostela, Spain}

\abstract{We present an explicit expression
for the topological invariants associated to $SU(2)$ monopoles
in the fundamental representation on  spin four-manifolds.
The computation of these invariants is based on the analysis of their
corresponding topological quantum field theory, and it turns
out that they can be expressed in terms of Seiberg-Witten invariants.
In this analysis we use recent exact results on the moduli
space of vacua of the untwisted $N=1$ and $N=2$ supersymmetric
counterparts of the topological quantum field theory under
consideration, as well as on electric-magnetic
duality for $N=2$ supersymmetric gauge theories.}

\endpage
\pagenumber=1

\chapter{Introduction.} Recently, there has been a great progress in the
understanding of the non-perturbative aspects of $N=1$ \REF\seiberg{N.
Seiberg \journal\pl&B318(93)469}  \REF\vacua{N. Seiberg
\journal\pr&D49(94)6857} \REF\superp{K. Intriligator, R.G. Leigh and N.
Seiberg \journal\pr&D50(94)1052} \REF\intin{K. Intriligator
\journal\pl&B336(94)409} \REF\fases{K. Intriligator and N. Seiberg
\journal\np&B431(94)551}  [\seiberg -\fases] and $N=2$ \REF\wspure{N.
Seiberg and E. Witten \journal\np&B426(94)19} \REF\wsmatter{N. Seiberg  and
E. Witten\journal\np&B431(94)484} [\wspure, \wsmatter] supersymmetric gauge
theories in four dimensions. On the one hand, holomorphy constraints and
non-perturbative non-renormalization  theorems have allowed to obtain exact
results for the behavior of the $N=1$ superpotentials present  in a wide
class of models. On the other hand, exact results on the quantum moduli
space of vacua  and on the low-energy effective actions of $N=2$
supersymmetric Yang-Mills theory and $N=2$ supersymmetric QCD have  been
obtained. These achievements have provided
 an explicit realization of electric-magnetic duality.

One of the most remarkable applications  of the exact solution of $N=2$ pure
Yang-Mills theory has been the reformulation in \REF\mfm{E.
Witten\journal\mrl&1(94)769} [\mfm] of Donaldson theory \REF\donfirst{S.K.
Donaldson\journal\jdg&18(83)279} \REF\don{S.K.
Donaldson\journal\topo&29(90)257} \REF\donkron{S.K. Donaldson and P.B.
Kronheimer,  {\it The Geometry Of Four-Manifolds}, Oxford Mathematical
Monographs, 1990}  [\donfirst, \don,\donkron] with gauge group $SU(2)$. It
is by now well-known \REF\tqft{E. Witten\journal\cmp&117(88)353}  [\tqft]
that Donaldson theory can be formulated as a certain twisted version of
$N=2$ Yang-Mills theory.  Using electric-magnetic duality  of this model one
can obtain an equivalent theory which involves an abelian  connection
coupled to matter in a pair of monopole equations. The new moduli problem is
much  more tractable than the original one, and it turns out that the
Donaldson polynomial invariants [\don] can be expressed in terms of certain
topological invariants associated to the abelian theory and called
Seiberg-Witten  invariants. The topological quantum field theory associated
to this new moduli space has been constructed in \REF\toplag{J.M.F.
Labastida and M. Mari\~no\journal\pl&B351(95)146} [\toplag] using the
Mathai-Quillen formalism.  Donaldson invariants  for K\"ahler manifolds were
computed previously by Witten in \REF\wjmp{E. Witten\journal\jmp&35(94)5101}
[\wjmp]. He showed that on a K\"ahler manifold it is possible to obtain a
topological symmetry for the $N=2$ Yang Mills theory which comes from an
$N=1$ subalgebra in such a way that the topological character is preserved
after perturbing the original  theory with an $N=1$ supersymmetric mass
term. The resulting $N=1$ theory reduces  at low energies to the $N=1$ pure
Yang-Mills theory and therefore one can use the results about its vacuum
structure  \REF\witsusy{E. Witten \journal\np&B202(82)253} [\witsusy] to
compute the correlation functions. The same procedure has been applied in
\REF\wv{C. Vafa and E. Witten \journal\np&B431(94)3}[\wv] to compute the
partition function of $N=4$ Yang-Mills theory on a K\"ahler manifold.

The monopole equations proposed in  [\mfm] have a natural non-abelian
generalization which appears in topological quantum field theories involving
the  minimal coupling of Donaldson-Witten theory to a twisted $N=2$ matter
multiplet.  These theories were constructed in \REF\rok{A. Karlhede and M.
Ro\v cek\journal\pl&B212(88)51} \REF\alab{M. Alvarez  and J.M.F.
Labastida\journal\pl&B315(93)251}   \REF\alabas{M. Alvarez and J.M.F.
Labastida\journal\np&B437(95)356} [\rok,\alab,\alabas], and related, more
general topological quantum field  theories have been analyzed in
\REF\ans{D. Anselmi and P. Fr\'e\journal\np&B392(93)401
\journal\np&B404(93)288\journal\np&B416(94)25\journal\pl&B347(95)247;
M. Bill\'o, R. D'Auria, S. Ferrara, P. Fr\'e, P. Soriani and A. van Proeyen,
``R-Symmetry and the Topological Twist of N=2 Effective Supergravities of
Heterotic Strings", hep-th/9505123} [\ans].  The non-abelian monopole
equations were studied in \REF\nonab{J.M.F. Labastida  and M.
Mari\~no\journal\np&B448(95)373} [\nonab] as a generalization of Donaldson
theory on four-manifolds, and the corresponding topological quantum field
theory was constructed in geometrical terms using the Mathai-Quillen
formalism. \REF\park{S. Hyun, J. Park and J.S. Park, ``Topological QCD",
hep-th/9503201} Other studies of these equations can be found in
[\ans,\park]. $U(N)$ monopole equations  have been considered from a
mathematical point of view in  \REF\oko{Ch. Okonek and A. Teleman,  ``The
Coupled Seiberg-Witten Equations, Vortices, And Moduli Spaces of Stable
Pairs",  to appear in {\it Int. J. Math.}} \REF\okodos{Ch. Okonek and A.
Teleman, ``Quaternionic Monopoles", alg-geom/9505029}[\oko, \okodos],  where
their relation to vortex equations on K\"ahler manifolds  \REF\brad{S.B.
Bradlow \journal\cmp&135(90)1 \journal\jdg&33(91)169} \REF\oscar{O. Garc\'\i
a-Prada \journal\cmp&156(93)527 \journal\ijm&5(94)1} [\brad, \oscar] is
stressed.

The aim of the present paper is to compute the topological correlation
functions of the topological field theory associated to $SU(2)$ monopoles in
the fundamental representation of the gauge group. This gives, as in the
Donaldson-Witten case, topological invariants which are polynomials in the
two-dimensional and four-dimensional cohomology of the moduli space. The
strategy of the computation is the following. First we will show that the
topological field theory introduced in [\nonab] is equivalent to a twisted
$N=2$ supersymmetric Yang-Mills theory coupled to one matter hypermultiplet,
and that on a K\"ahler manifold one can obtain a topological symmetry coming
from a $N=1$ subalgebra, extending in this way the result of [\wjmp] for
pure $N=2$ supersymmetric Yang-Mills. This makes possible to compute in the
$N=1$ theory obtained after perturbing with a mass term. The vacuum
structure of the resulting theory is obtained using the low-energy
description of the $N=2$ theory in [\wsmatter]. The computation of the
polynomial invariants is performed on a K\"ahler manifold following the
procedure in [\wjmp]. Then we will use electric-magnetic duality of the
$N=2$ theory and the results of [\mfm] to obtain a general expression for
spin manifolds. This expression can be written in terms of Seiberg-Witten
invariants, as one should guess from the analysis in [\wsmatter]. Therefore,
Seiberg-Witten invariants  seem to underlie the moduli space of
anti-self-dual (ASD) $SU(2)$ instantons as well as the moduli space of
$SU(2)$ monopoles.

The paper is organized as follows. In sect. 2 we review the
construcion of [\nonab] and we relate it to the standard
topological twisting of
$N=2$ supersymmetric QCD. In addition, we
present the observables of the theory, we
perform the twist on a K\"ahler manifold, and formulate the
perturbed theory. In sect. 3 the
vacuum structure of the $N=2$ and $N=1$ theory is analyzed
and we obtain the symmetry patterns needed in our
computations. In sect. 4 we compute the polynomial
invariants, first from the $N=1$ point of view on
a K\"ahler manifold, and then using electric-magnetic
duality and the low-energy structure of the $N=2$ theory
on a general spin manifold. In sect. 5 we consider the
twisted $N=2$ supersymmetric
QCD theory with a massive hypermultiplet,
and we obtain the vacua of the perturbed $N=1$ supersymmetric
theory in
order to support the previous analysis. In sect.
6 we state our conclusions and prospects for future work. The first
appendix contains some observations about the parity symmetry
of the $N=2$ theory. Finally, in the second appendix we rederive
the results about the vacuum structure of the $N=1$ theory
from its exact superpotential.

\endpage

\chapter{Non-abelian monopoles}

\def\mna{{{\cal M}_{\hbox{\sevenrm NA}}}}
\def\masd{{{\cal M}_{\hbox{\sevenrm ASD}}}}
\def\ex{{\hbox{\rm e}}}

In this section we will make a  brief presentation
 of the non-abelian monopole equations and their corresponding
topological quantum field theory. We refer the reader
to [\nonab] for the details of the construction.
In addition, we will discuss how this theory
can be regarded as a twisted version of $N=2$ non-abelian
Yang-Mills theory coupled to a massless $N=2$ matter hypermultiplet.
Finally we will consider the theory on a K\"ahler manifold, and we will
show how in this case the theory can be perturbed generating an $N=1$
supersymmetric mass term
while preserving the topological character of the theory.

\section{Non-abelian monopole equations}

Let $X$ be an oriented, compact, spin four-manifold endowed with a
Riemannian
structure given by a metric $g$. We will restrict our analysis to spin
manifolds since the arguments used in the following sections are only valid
for this type of manifolds. The generalization of the  non-abelian monopole
equations for other manifolds can be done using a Spin$_c$ structure. Work
in this direction has appeared recently  [\oko,\park]. The positive and
negative chirality spin bundles of $X$ will be denoted by $S^+$ and $S^-$,
respectively. Let $P$ be a principal fibre bundle with some compact,
connected, simple group $G$, and let $E$ be an associated vector bundle to
the principal bundle $P$ via a representation $R$ of $G$. The Lie-algebra
associated to $G$ will be denoted by ${\bf g}$. Given  this data we will
consider the field space  ${\cal M} = {\cal A} \times \Gamma(X,S^+\otimes
E)$, where  ${\cal A}$ is the space of $G$-connections on $E$,  and
$\Gamma(X,S^+\otimes E)$ is the space of positive-chirality spinors taking
values in this representation space. The group ${\cal G}$ of gauge
transformations of the  bundle $E$ acts locally on the elements of
${\cal M}$
in the following way: $$
\eqalign{ &g^{*}(A_\mu)=-igd_\mu g^{-1}+gA_\mu g^{-1},\cr
&g^{*}(M_{\alpha})=g M_{\alpha},\cr }
\eqn\gauge
$$
where $A\in {\cal A}$,
$M \in \Gamma (X, S^{+} \otimes E)$, and $g$ takes values
in the representation $R$ of the group $G$. Notice that in
\gauge\ while $\mu$ is
a space-time index, $\alpha$ is a positive-chirality spinor
index. In this paper we use the same notation as in [\nonab] where
the index conventions are described in detail.
In terms of the covariant
derivative $d_A=d+i[A,\;\;]$, the infinitesimal form of the transformations
\gauge\ can be considered as a linear operator:
$$
C(\phi)=(-d_{A}\phi ,i\phi M) \in \Omega^{1}(X,{\bf g}_E) \oplus \Gamma
(X, S^{+}
\otimes E),\,\,\,\,\ \phi \in \Omega^{0}(X,{\bf g}_E).
\eqn\flor
$$ being $\phi$ such
that $g=\exp(i\phi)$.
Let us consider a trivial vector bundle ${\cal V}$ over ${\cal M}$ with
fibre ${\cal F} =
\Omega^{2,+}(X,{\bf g}_E)\oplus\Gamma(X,S^-\otimes E)$, where
the self-dual differential forms take values in the Lie-algebra
representation, ${\bf g}_E$, associated to $R$. The non-abelian
monopole equations define a moduli space which is the zero locus
of a section on this bundle, $s:{\cal M} \longrightarrow {\cal V}$.
Actually, due to the presence of the gauge symmetry \gauge\
one must account for the action of the gauge group ${\cal G}$ in
both, ${\cal M}$ and ${\cal V}$. One must therefore consider
the associated section $\hat s: {\cal M}/{\cal G} \longrightarrow
{\cal V}/{\cal G}$. The resulting moduli space will be denoted
by $\mna$.

We will restrict ourselves to the case $G=SU(N)$ and $R$ its fundamental
representation, $R={\bf N}$. The generalization
of the monopole equations to other simple gauge groups
and other representations is straightforward.
The non-abelian monopole equations take the form [\nonab]:
$$
\eqalign{ &F_{\alpha\beta}^{+ij}+{i \over 2}({\overline
M}^j_{(\alpha} M^i_{\beta )}-{\delta^{ij} \over N}{\overline M}^k_{(\alpha}
M^k_{\beta )})=0, \cr
&(D_{\alpha\dot\alpha}M^\alpha)^i=0,\cr}
\eqn\nonabmon
$$
where $F_{\alpha\beta}^{+ij}$ are in the fundamental
  representation,
\ie, $F_{\alpha\beta}^{+ij}=F_{\alpha \beta}^{+a} (T^a)^{ij}$,
being $T^a$, $a=1,\dots,N^2-1$,
the generators of the Lie algebra in the  representation ${\bf N}$.
In the first equation of \nonabmon\ (and similar ones in this paper), a
sum in the repeated index $k$ is understood. The second equation
in \nonabmon\ is simply the Dirac equation with the Dirac operator
coupled to the gauge connection in the fundamental representation.

The section of the bundle ${\cal V}$, $s:{\cal M}\longrightarrow
{\cal V}$, corresponding to \nonabmon\ has the following form:
$$
s(A,M)=\Big({a \over \sqrt 2} \big(F^{+ ij}_{\alpha \beta}+{i\over 2}
({\overline M}_{(\alpha}^j  M_{\beta )}^i-{\delta^{ij}\over d_R} {\overline
M}_{(\alpha}^k  M_{\beta )}^k)\big), (D_{\alpha {\dot \alpha}
}M^{\alpha})^i\Big),
\eqn\lasection
$$
where $a$ is complex number different from zero. This number $a$ was taken
to be one in [\nonab] because then certain useful vanishing
theorems can be utilized, as first shown in [\wjmp] for
the abelian case. As it will be dicussed below, the
observables of the topological quantum field theory
associated to the section \lasection\ are independent
of the value chosen for $a$ as long as $a\not= 0$.

Some aspects of the moduli space of solutions of the non-abelian
monopole equations \nonabmon\ modulo gauge transformations have been
studied in [\nonab]. In particular, the virtual dimension of this
moduli space, ${\rm dim}\, \mna$, turns out to be:
$$
{\rm dim} \, \mna ={\rm dim} \, \masd +2\,{\rm
index}\, D
=(4N-2)c_2(E)-{N^2-1 \over 2}(\chi +\sigma)-{d_{R} \over 4}\sigma,
\eqn\dim
$$
where $\chi$ and $\sigma$ are the Euler characteristic
and the signature, respectively,  of the manifold $X$,
and $c_2(E)$ is the second Chern class of the representation bundle and
equals the instanton number $k$.
In \dim\ $\masd$ denotes the moduli space of anti-self-dual
(ASD) instantons and ${\rm dim}\, \masd$ its virtual dimension,
$$
{\rm dim}\,
\masd = 4N\,c_2(E)-(N^2-1)(\chi+\sigma)/2,
\eqn\masd
$$
which is the index of
the ASD complex:
$$
0 \too \Omega^{0}(X,{\bf g}_E)
\buildrel d_A \over \too \Omega^{1}(X,{\bf g}_E)
\buildrel p^{+}d_A \over \too
 \Omega^{2,+} (X,{\bf g}_E)  \too 0.
\eqn\asd
$$
${\rm index}\, D$ denotes the index of the Dirac operator coupled to the
connection on $E$ , which is given by:
$${\rm index}\, D=\int_X {\rm ch}(E) {\hat
A}(X)=-{N\over 8}\sigma -c_2(E).
\eqn\dirac$$
Notice that on a four-dimensional
spin manifold the index of the Dirac complex is
given by $-\sigma /8$, and is always an integer.
Therefore $\sigma \equiv 0 \,\
{\rm mod}\,\ 8$. Also notice that on a four-dimensional K\"ahler manifold
$$
\chi+\sigma=2-2b_1+b_2^{+}=4(1-h^{1,0}+h^{2,0}),
\eqn\laura
$$
where $b_1$ is the first Betti number, $b_2^+$ is the dimension
of $H^{2,+}(X)$, and $h^{1,0}$, $h^{2,0}$
denote Hodge numbers. Therefore, on a
K\"ahler manifold, the quantity
$$
\Delta={ \chi+\sigma \over 4},
\eqn\ladelta
$$
is always an integer.

When dealing with moduli spaces associated to the solutions of certain
equations, as it happens in our case, one must require certain conditions in
order to have a well defined moduli problem. These requirements concern the
orientability of the moduli space (which is equivalent to require that the
corresponding topological field theory does not have global anomalies) and
the free action of the group of gauge transformations on the space of
solutions. As it was argued in [\nonab], these conditions are fulfilled in
the non-abelian monopole problem as long as they are fulfilled in the
Donaldson theory with the same gauge group. In our case we are concerned
with $SU(2)$, and the corresponding conditions reduce to $b_2^+ >1$.  In the
following we will suppose that this condition holds for our four-dimensional
spin manifold $X$. On a K\"ahler manifold, $b_2^+=2h^{2,0}+1$ and the above
condition is equivalent to $H^{2,0}(X) \not= 0$.

In [\nonab] a preliminary analysis of the moduli space of solutions of the
$SU(N)$ monopole equations on compact K\"ahler manifolds was done. This
moduli space has three branches: one of them corresponds to $M=0$ and is
the moduli space of ASD instantons of Donaldson theory, which is thus
contained in ${\rm dim} \, \mna$. The second  branch corresponds to pairs
consisting of an equivalence class of holomorphic $Sl(N,{\bf C})$ bundles
together with a holomorphic section of $K^{1/2}\otimes E$ modulo
$Sl(N,{\bf C})$ gauge transformations. The third branch is similar to the
second branch, but we must  consider instead holomorphic sections of
$K^{1/2}\otimes {\overline E}$. In addition we need some stability
conditions for  the pair in order to guarantee the existence of solutions.
These stability conditions have an  algebraico-geometric character and
appear in the Hermite-Einstein equations  \REF\donstab{S.K. Donaldson
\journal\plms&53(85)1} [\donstab], in the Hitchin equations on  Riemann
surfaces \REF\hitchin{N.J. Hitchin \journal\plms&55(87)59} and in the
vortex equations [\brad, \oscar]. The  non-abelian monopole equations on a
K\"ahler manifold  with gauge group $SU(N)$ are closely  related to vortex
equations where, as we are taking the tensor product  of the original
bundle $E$ or its conjugate with $K^{1/2}$, the resulting bundle has a
fixed
 determinant. This situation has been analyzed in [\okodos] where the
corresponding stability condition has been obtained. Further work on the
relation between non-abelian monopole equations  and the vortex equations
will appear elsewhere  \REF\nos{O. Garc\'\i a -Prada, J.M.F. Labastida and
M. Mari\~no, to appear} [\nos].

The topological action corresponding to the moduli problem
leading to $\mna$ was constructed in [\nonab] using the
Mathai-Quillen formalism. In order to present the form of
this action we need to introduce first a variety of fields.
Let $(\psi,\mu)$ be an element of the tangent space to the moduli
space ${\cal M}$ at the point $(A,M)$,
$(\psi,\mu) \in T_{(A,M)}{\cal M}=T_{A}{\cal A}
\oplus T_{M}\Gamma (X, S^{+} \otimes E)=\Omega^{1}(X,{\bf g}_E)
\oplus \Gamma (X, S^{+} \otimes E)$, and let $\phi$ be an element of
$\Omega^0(X,{\bf g}_E)$, $\phi\in \Omega^0(X,{\bf g}_E)$.
The fields $(A,M)$, $(\psi,\mu)$ and $\phi$ have ghost numbers
0, 1 and 2, respectively.
Associated to the fiber  ${\cal F}$ we introduce
fields $(\chi,v_{\dot\alpha})\in
{\cal F} =  \Omega^{2,+}(X,{\bf g}_E)\oplus\Gamma(X,S^-\otimes E)$,
with ghost number $-1$. In addition, fields $\lambda$ and $\eta$ in
$\Omega^0(X,{\bf g}_E)$ with ghost number $-2$, and $-1$, respectively,
as well as an  auxiliary sector made out of ghost-numer zero fields
$(H,h)\in \Omega^{2,+}(X,{\bf g}_E)\oplus\Gamma(X,S^-\otimes E)$
are introduced. All fields with even ghost number are commuting while the
ones with odd ghost number are anticommuting.

The topological action can be written very simply with the help
of the BRST symmetry present in the formalism. Under this symmetry
the fields of the theory transform in the following way:
$$
\eqalign{ &[Q,A]=\psi,\cr
&\{Q,\psi \}=d_A\phi,\cr
&[Q, \phi]=0,\cr
&\{Q,\chi_{\mu\nu} \}=H_{\mu\nu}, \cr
&[Q,H_{\mu\nu}]=i[\chi_{\mu\nu},\phi], \cr
&[Q, \lambda]=\eta, \cr}
\qquad\qquad
\eqalign{
&[Q,M_{\alpha}^i]=\mu_{\alpha}^i, \cr
&\{Q, \mu_{\alpha}^i \}=-i\phi^{ij} M_{\alpha}^j,\cr
& \{Q, v_{\dot \alpha}^i \}=h_{\dot \alpha}^i, \cr
&[Q,h_{\dot \alpha}^i]=-i \phi^{ij} v_{\dot \alpha}^j, \cr
& \{Q, \eta \}=i[\lambda,\phi].\cr}
\eqn\pera
$$
This symmetry closes up to a gauge transformation \gauge\ generated
by the field $-\phi$. As anticipated, the action is written as
a $Q$-exact expression:
$$
\eqalign{S=&
\Big\{ Q, \int _{X} e \Big[i\chi^{\alpha \beta ji}\Big({a \over {\sqrt 2}}
\big(F^{+ij}_{\alpha \beta} +{i \over 2}({\overline M}_{(\alpha}^j
M_{\beta)}^i -{\delta^{ij}\over d_R}{\overline M}_{(\alpha}^k
M_{\beta)}^k)\big) - {i \over 4} H_{\alpha \beta} \Big)\cr
&\,\,\,\,\,\,\,\,\,\,\,\,\,\,\,\,\,\,\,\,\,\,\,\,\,
 +{i\over 2} ({\bar v}^{\dot \alpha}D_{\alpha \dot \alpha}
M^{\alpha}+{\overline M}^{\alpha}D_{\alpha \dot \alpha}v^{\dot \alpha})
+{1\over 8} ({\bar v}^{\dot \alpha} h_{\dot \alpha}-{\bar h}_{\dot
\alpha}v^{\dot \alpha} ) \cr
&\,\,\,\,\,\,\,\,\,\,\,\,\,\,\,\,\,\,\,\,\,\,\,\,\,
+ i\tr(\lambda \wedge *d_A^{*} \psi) + {1 \over 2}   ({\bar
\mu}^{\alpha}\lambda M_{\alpha}- {\overline M}^{\alpha} \lambda
\mu_{\alpha})   \Big] \Big\}= S_0+S_M \cr}
\eqn\curva
$$
where,
$$
\eqalign{
S_0 = & \int _{X}\Big\{ e \Big[{i a \over {\sqrt 2}}
 H^{\alpha \beta ji} F^{+ij}_{\alpha \beta}
-{ia \over {\sqrt 2}}\tr\big(\chi ^{\alpha
\beta}(p^+(d_A\psi))_{\alpha\beta}\big) \cr &
\,\,\,\,\,\,\,\,\,\,\,\,\,\,\,\,\,\,\,\,\,\,\,\,\,\,\,\,\,
\,\,\,\,\,\,\,\,\,\,\,
+{1 \over 4} \tr(H^{\alpha \beta} H_{\alpha \beta})
-{i\over 4} \tr(\chi^{\alpha\beta}
[\chi_{\alpha\beta},\phi])\Big]  \cr &
\,\,\,\,\,\,\,\,\,\,\,\,\,\,\,\,
-\tr\big(- i\eta \wedge *d_A^{*} \psi
-i \lambda \wedge *d_A^{*}d_A \phi
- \lambda\wedge *[*\psi,\psi] \big)\Big\} \cr}
\eqn\isabel
$$
and,
$$
\eqalign{
S_M=  & \int _{X}e \Big[{- a \over 2 {\sqrt 2}}
H^{\alpha \beta ji} ({\overline
M}_{(\alpha}^j M_{\beta)}^i -{\delta^{ij}\over N}{\overline M}_{(\alpha}^k
M_{\beta)}^k) \cr  &\,\,\,\,\,\,\,\,\,\,\,\,\,\,
-{a\over \sqrt{2}}({\bar \mu}_{\alpha}\chi^{\alpha\beta} M
_{\beta}- {\overline M}_{\alpha}\chi^{\alpha\beta}\mu _{\beta})
+{i\over 2} ({\bar h}^{\dot \alpha}D_{\alpha \dot \alpha}
M^{\alpha}+{\overline M}^{\alpha}D_{\alpha \dot \alpha}h^{\dot \alpha})
 \cr
&\,\,\,\,\,\,\,\,\,\,\,\,\,
-{i \over 2}({\bar v}^{\dot \alpha}D_{\alpha \dot \alpha} \mu^{\alpha}-{\bar
\mu}^{\alpha}D_{\alpha \dot \alpha}v^{\dot \alpha}) - {1 \over 2}
 ({\overline M}^{\alpha} \psi_{\alpha \dot
\alpha}v^{\dot \alpha}- {\bar v}^{\dot \alpha} \psi_{\alpha \dot
\alpha}M^{\alpha})\cr
&\,\,\,\,\,\,\,\,\,\,\,\,\,\,
- {1 \over 2}   ({\bar \mu}^{\alpha}\eta M_{\alpha} + {\overline M
}^{\alpha} \eta \mu_{\alpha}) -   ({\bar
\mu}^{\alpha}\lambda\mu_{\alpha} -{i\over 2} {\overline M}^{\alpha}
\{\phi,\lambda\} M_{\alpha})\cr
&\,\,\,\,\,\,\,\,\,\,\,\,\,\,
+{1 \over 4} ({\bar h}^{\dot\alpha} h_{\dot\alpha}
+i{\bar v}^{\dot\alpha} \phi v_{\dot\alpha})\Big].\cr}
\eqn\manza
$$
The action $S_0$ is the one corresponding to Donaldson-Witten
theory. Its observables lead
to standard Donaldson invariants. The action $S_M$
contains  the `matter fields' and their couplings
to Donaldson-Witten theory. The observables associated
to the total action $S$ lead to new topological invariants.
Notice that the coefficient $a$ enters in \curva\
multiplying a $Q$-exact
term. This means that any variation in $a$ is $Q$-exact and therefore,
using standard arguments [\tqft], the observables of the theory
are independent of $a$ as long as $a\not= 0$.
In the rest of this subsection  we will take $a=1$.

The action $S_0$ differs from the one in [\tqft] by a term
of the form:
$$
\big\{ Q,  \int_X e \tr( \lambda [ \eta, \phi])\big\}.
\eqn\diferencia
$$
This term appears naturally when Donaldson-Witten theory
is regarded as a twist of $N=2$ supersymmetry, but its
presence is rather unnatural from the point of view of
the Mathai-Quillen formalism. However, being a $Q$-exact term involving
products of fields, the observables of the theory do not
depend on it.

After integrating out the  auxiliary fields
$H_{\alpha\beta}$ and $h_{\dot\alpha}$ in the action \curva\
one finds the on-shell action given in [\nonab]
(recall that we set $a=1$):
$$
\eqalign{\tilde S=&
\int_{X} e \big[ g^{\mu\nu}D_\mu\overline M^\alpha D_\nu
M_\alpha + {1\over 4} R \overline M^\alpha M_\alpha +{1\over 2}
\tr (F^{+\alpha\beta} F_{\alpha\beta}^+)
-{1\over 8}(\overline
M^{(\alpha} T^a M^{\beta)}) (\overline M_{(\alpha} T^a M_{\beta)})]
\cr
+&\int_{X} \tr\big(\eta \wedge *d_A^{*} \psi
-{i \over {\sqrt 2}}\chi ^{\alpha
\beta}(p^+(d_A\psi))_{\alpha\beta}
-{i\over 4}\chi^{\alpha\beta}[\chi_{\alpha\beta},\phi]
+ i \lambda \wedge *d_A^{*}d_A \phi + \lambda\wedge *[*\psi,\psi] \big)
\cr
+&\int_{X}e\Big(-i{\overline M}^{\alpha}\{\phi, \lambda\} M_{\alpha}  +{1
\over {\sqrt2}} ({\overline M}_{\alpha} {\chi}^{\alpha \beta} \mu _{\beta
}-{\bar \mu}_{\alpha}{\chi}^{\alpha \beta} M_{\beta}) -{i\over 2} ({\bar
v}^{\dot \alpha} D_{\alpha {\dot\alpha}} {\mu}^{\alpha}-{\bar \mu}^{\alpha}
D_{\alpha {\dot\alpha}}v^{\dot \alpha})  \cr
&\,\,\,\,\,\,\,\,\,\,\,\,\,\,\,\,\,\,-{1 \over 2}({\overline
M}^{\alpha}{\psi}_{\alpha {\dot \alpha}}v^{\dot \alpha}-{\bar v}^{\dot
\alpha}{\psi}_{\alpha {\dot \alpha}}M^{\alpha}) -{1\over2}  ({\bar
\mu}^{\alpha}  \eta M_{\alpha}+{\overline M}^{\alpha}
 \eta\mu_{\alpha}) +{i \over 4}{\bar v}^{\dot\alpha} \phi v_{\dot\alpha}
-{\bar \mu}^{\alpha} \lambda \mu _{\alpha}\Big).\cr}
\eqn\action
$$

\section{Observables}

The observables of the theory are those operators
in the cohomology of $Q$. From the transformations \pera\ follow
that the observables in the ordinary $SU(2)$ Donaldson-Witten theory
are also observables in this theory.
These observables are based on the $k$-forms operators
${\cal O}^{(k)}$ in [\tqft]:
$$
\eqalign{
{\cal O}^{(0)} =& -{1\over 4} \tr (\phi^2),\cr
{\cal O}^{(1)} =& -{1\over 2} \tr (\phi\psi), \cr
{\cal O}^{(2)} =& {1\over 2} \tr (i\phi F - {1\over 2}\psi\wedge\psi),\cr}
\qquad
\eqalign{
{\cal O}^{(3)} =& {i\over 2} \tr (\psi\wedge F),\cr
{\cal O}^{(4)} =& {1\over 2} \tr (F\wedge F).\cr}
\eqn\obser
$$
These operators satisfy the descent equations,
$$
d {\cal O}^{(k)} = \{ Q, {\cal O}^{(k+1)}\}.
\eqn\descent
$$
{}From these equations follow that if $\Sigma$ is a $k$-dimensional
homology cycle, then
$$
I(\Sigma) = \int_\Sigma {\cal O}^{(k)},
\eqn\moreobser
$$
is in the cohomology of $Q$. For simply connected four-manifolds, which is
the case of interest in this paper, $k$-dimensional homology cycles only
exist for $k=0,2,4$. For $k=4$ the cycle $\Sigma$ is the four-manifold $X$
and $I(X)$ is the instanton number.

We have not found any new invariant involving matter fields. This problem
should be addressed from the point of view of the universal instanton, but
presumably the absence of matter invariants means that the universal bundle
associated to the non-abelian monopole equations is the pullback of the
universal bundle associated to Donaldson theory.  For simplicity we will
denote the observable corresponding to $k=0$ by ${\cal O}(x)$. The most
general observable which we will consider in this paper will have the form:
$$ {\cal O}(x_1)\cdots{\cal O}(x_r) I(\Sigma_1)\cdots I(\Sigma_s).
\eqn\general $$
Correlation functions involving operators of the form \general\ vanish
unless the following selection rule holds: $$
4r+2s={\rm dim} \, \mna.
\eqn\selrule$$
where ${\rm dim} \, \mna$ is given in \dim\ with  $N=2$. These
 correlation functions of the topological field theory are interpreted
mathematically as intersection forms in the moduli space. The operator $\cal
O$ represents a cohomology class of degree four, and $ I(\Sigma)$ represents a
cohomology class of degree two. The condition \selrule\ simply says that the
integral of these differential forms vanishes unless the total degree equals
the dimension of the moduli space. This has a natural interpretation in
field-theoretical terms [\tqft].
The dimension of the moduli space corresponds
to the index of the operator:
$$
T=ds \oplus C^{\dagger}: \,\
\Omega^{1}(X,{\bf g}_E) \oplus \Gamma (X, S^{+} \otimes
E) \too
 \Omega^{0}(X,{\bf g}_E) \oplus \Omega^{2,+} (X,{\bf g}_E) \oplus \Gamma (X,
S^{-} \otimes E),
   \eqn\file
$$
which gives the instanton deformation complex for the moduli problem of
non-abelian monopoles. But this is also the operator associated to the
grassmannian fields in \action, and its index gives the anomaly in the
ghost number. The selection rule \selrule\ is therefore the 't Hooft rule
which says that fermionic zero modes in the path integral measure should be
soaked up in the correlation functions.

As it is usual in quantum field theory,
we will group all the correlation functions of operators like \general\ in a
generating function,
$$
\langle {\rm exp} (\sum_{a}\alpha_{a}I(\Sigma_{a})+\mu
{\cal O}) \rangle,
\eqn\generatriz
$$
summed over instanton numbers of the bundle
$E$. In \generatriz\ the $\Sigma_{a}$ denote a basis of the two-dimensional
homology of $X$, and therefore $a=1, \cdots, {\rm dim} \,\ H_2(X,{\bf Z})$.
\section{Twist of the $N=2$ supersymmetric theory}

The action  \action\ can be obtained from the twist of $N=2$
supersymmetric Yang-Mills theory with gauge group $SU(N)$
coupled to an $N=2$ hypermultiplet in the fundamental
representation
[\rok,\alab,\alabas,\ans,\park].
The basic idea involved in the twisting is the
following. In ${\bf R}^4$ the global symmetry group
of $N=2$ supersymmetry is ${\cal H}=SU(2)_L\otimes SU(2)_R\otimes
SU(2)_I\otimes U(1)_{\cal R}$,
where ${\cal K}=SU(2)_L\otimes SU(2)_R$ is the rotation group
and $SU(2)_I$ and $U(1)_{\cal R}$ are internal symmetry groups.
The supercharges
$Q_\alpha^i$ and $\overline Q_{\dot\alpha i}$ of $N=2$ supersymmetry
transform under ${\cal H}$ as $(1/2,0,1/2)^1$ and $(0,1/2,1/2)^{-1}$
respectively. The twist consists in considering as the
rotation group the group  ${\cal K}'=SU(2)_L'\otimes SU(2)_R$
where $SU(2)_L'$ is the diagonal subgroup of
$SU(2)_L\otimes SU(2)_I$. Under the new global symmetry group
${\cal H}'={\cal K}'\otimes U(1)_{\cal R}$
the supercharges transforms as
$(1/2,1/2)^{-1}\oplus(1,0)^1\oplus(0,0)^1$.
In the proccess of twisting the isospin index $i$ becomes
a spinor index, $Q_\alpha^i \rightarrow Q_\alpha{}^\beta$
and $Q_{\dot\alpha i} \rightarrow Q_{\dot\alpha\beta}$,
and the trace of $Q_\alpha{}^\beta$, $Q=Q_\alpha{}^\alpha$,
becomes a $(0,0)$ rotation invariant operator.

If there is no $N=2$ central extension, from the
supersymmetry algebra follows that $Q$ obeys $Q^2=0$.
This operator can be regarded as a BRST operator and the
$U(1)_{\cal R}$ charges as ghost numbers.
In ${\bf R}^4$ the original and the twisted theory are the
same.  For other manifolds the two theories are certainly
different since the stress tensor changes.
On the other hand, due to the fact that
the operator $Q$ is an scalar, it also exists
for  arbitrary four manifolds. The existence
of this operator is what gives the topological character
to twisted theories.

We will begin briefly describing the twist of $N=2$ supersymmetric
Yang-Mills theory. Then, we will consider the case of its coupling
to an $N=2$ hypermultiplet.
The field content of the minimal $N=2$ supersymmetric Yang-Mills theory
with gauge group $G$ is the following:
a gauge field $A_{\alpha\dot\alpha}$
(using the notation in [\nonab],
$A_{\alpha\dot\alpha}=e_\mu^m(\sigma_m)_{\alpha\dot\alpha}$),
fermions $\lambda_\alpha^i$
and $\overline\lambda_{\dot\alpha i}$, a complex scalar $B$, and an
auxiliary field $D_{ij}$ (symmetric in $i$ and $j$). All these fields
are considered in the adjoint representation of the gauge
group $G$. Under the twisting these fields become:
$$
\eqalign{
A_{\alpha\dot\alpha} \;\; (1/2,1/2,0)^0 \;\; & \longrightarrow  \cr
\lambda_\alpha^i \;\; (1/2,0,1/2)^{-1} \;\;&\longrightarrow  \cr
\overline\lambda_{\dot\alpha i} \;\; (0,1/2,1/2)^1 \;\;& \longrightarrow \cr
B \;\; (0,0,0)^{-2} \;\;& \longrightarrow \cr
B^\dagger \;\; (0,0,0) \;\;&\longrightarrow \cr
D_{ij} \;\; (0,0,1)^0 \;\;& \longrightarrow \cr}
\quad
\eqalign{
& A_{\alpha\dot\alpha} \;\; (1/2,1/2)^0, \cr
& \eta  \;\; (0,0)^{-1}, \;\;\; \chi_{\alpha\beta} \;\; (1,0)^{-1}, \cr
& \psi_{\alpha\dot\alpha} \;\; (1/2,1/2)^1, \cr
& \lambda \;\; (0,0)^{-2}, \cr
& \phi \;\; (0,0)^2, \cr
& H_{\alpha\beta} \;\; (1,0)^0, \cr}
\eqn\twist
$$
where we have indicated the quantum numbers carried out
by the fields relative to the group ${\cal H}$ before the twisting,
and to the group ${\cal H}'$ after the twisitng.
Notice that the fields $\chi_{\alpha\beta}$ and $H_{\alpha\beta}$
are symmetric in $\alpha$ and $\beta$ and therefore
they can be regarded as components of
 two self-dual two-forms. The definitions of the
twisted fields in terms of the untwisted ones are the obvious
ones from \twist. The only ones which need clarification are
the conventions taken for $\eta$ and $\chi_{12}$. Our choice is:
$$
\lambda^1_1={1\over 2} \eta - \chi_{12},
\;\;\;\;\;\;\;
\lambda^2_2={1\over 2} \eta + \chi_{12}.
\eqn\debbie
$$

The $Q$-transformations of the twisted fields can be obtained
very simply from the $N=2$ supersymmetry transformations. These
last transformations are generated by the operator
$\eta^\alpha_i Q_\alpha^i + \overline \eta^{\dot\alpha i}
\overline Q_{\dot\alpha i}$ where $\eta^\alpha_i$ and
$\overline \eta^{\dot\alpha i}$ are anticommuting parameters.
To get the $Q$-transformations of the fields one must
consider $\overline \eta^{\dot\alpha i}=0$ and replace
$\eta^\alpha_i \rightarrow \rho \delta^\alpha_\beta$, being
$\rho$ an arbitrary scalar anticommuting parameter.
The twisting leads to the transformations \pera\ for the
twisted fields on the right hand side of \twist, and to
a twisted action which turns out to be the action
$S_0$ in \isabel\ for some value of $a$ plus a term of the
form \diferencia.

Before discussing the coupling of an $N=2$ hypermultiplet
let us make a few comments on the twisting from a $N=1$
superspace point of view. In $N=1$ superspace only one
of the supersymmetries is manifest, and therefore the $N=1$
superfields do not have well defined quantum numbers
respect to the internal $SU(2)_I$ symmetry.
The $N=2$ supersymmetric multiplet
contains an $N=1$ vector multiplet and an $N=1$ chiral multiplet.
These multiplets are described in $N=1$ superspace in terms
of $N=1$ superfields $W_\alpha$ and $\Phi$ satisfying the
constraints $\overline D_{\dot\alpha} W_\alpha =0$,
$D^\alpha W_\alpha + \overline D^{\dot\alpha} \overline
W_{\dot\alpha}=0$ and $\overline D_{\dot\alpha} \Phi =0$,
where $D_\alpha$ and $\overline D_{\dot\alpha}$ are
$N=1$ superspace covariant derivatives (we use the conventions
in  \REF\superb{S.J. Gates, M.T. Grisaru, M. Ro\v cek and W. Siegel,
``Superspace", Benjamin, 1983}
[\superb]).
The $N=1$ superfields $W_\alpha$ and $\Phi$ have $U(1)_{\cal R}$
charges $-1$ and $-2$ respectively.
The component
fields of the $N=1$ superfields $W_\alpha$  and $\Phi$ are:
$$
\eqalign{
W_\alpha, \;\; \overline W_{\dot\alpha}
 \;\; & \longrightarrow \;\;
A_{\alpha\dot\alpha}, \;\;  \lambda^1_\alpha, \;\;
\overline\lambda_{\dot\alpha 1}, \;\; D_{12}, \cr
\Phi, \;\; \Phi^\dagger \;\;& \longrightarrow \;\;  B, \;\;
\lambda^2_\alpha,  \;\; D_{11},  \;\; B^\dagger, \;\;
\overline\lambda_{\dot\alpha 2}, \;\; D_{22}.\cr}
\eqn\trapecio
$$
The $U(1)_{\cal R}$ transformations of the $N=1$ superfields are:
$$
W_\alpha \rightarrow \ex^{-i\phi} W_{\alpha}(\ex^{i\phi}\theta),
\;\;\; {\rm and} \;\;\;
\Phi \rightarrow \ex^{-2i\phi} \Phi(\ex^{i\phi}\theta).
\eqn\salto
$$
Notice that these transformations
are consistent with the assignment in \twist.

In $N=1$ superspace the action of $N=2$ supersymmetric Yang-Mills theory
takes the form:
$$
\int d^4x \, d^2\theta d^2\overline\theta\,\Phi^\dagger \ex^V \Phi+
\int d^4x \, d^2\theta \, \tr (W^\alpha W_\alpha) +
\int d^4x \, d^2\overline\theta \, \tr (\overline  W^{\dot\alpha}
\overline W_{\dot\alpha}),
\eqn\superespacio
$$
where $V$ is the vector superpotential.
An important feature of this action is that due to the
constraint $D^\alpha W_\alpha + \overline D^{\dot\alpha} \overline
W_{\dot\alpha}=0$ the last two terms in \superespacio\ differ by
a term which is proportional to the second Chern class.

$N=2$ matter is usually represented by  $N=2$ hypermultiplets.
The hypermultiplet contains a complex scalar isodoublet
$q^i$, fermions $\psi_{q \alpha}$, $\psi_{\tilde q \alpha}$,
$\overline \psi_{q \dot\alpha}$, $\overline \psi_{\tilde q
\dot\alpha}$, and a complex scalar isodoublet auxiliary field
$F^i$. The fields
$q^i$, $\psi_{q \alpha}$, $\overline \psi_{\tilde q
\dot\alpha}$, and $F^i$ are in the fundamental representation
of the gauge group, while the fields
$q_i^\dagger$, $\psi_{\tilde q \alpha}$,
$\overline \psi_{q \dot\alpha}$, and $F_i^\dagger$
are in the conjugate representation. Under the twisting
these fields become:
$$
\eqalign{
q^i\;\;  (0,0,1/2)^0 \;\; & \longrightarrow \cr
\psi_{q \alpha} \;\; (1/2,0,0)^1 \;\; & \longrightarrow \cr
\overline \psi_{\tilde q \dot\alpha} \;\; (0,1/2,0)^{-1} \;\;
& \longrightarrow \cr
F^i \;\; (0,0,1/2)^2 \;\; & \longrightarrow \cr
q_i^\dagger \;\;  (0,0,1/2)^0 \;\; & \longrightarrow \cr
\overline \psi_{q \dot\alpha} \;\; (0,1/2,0)^{-1} \;\;
& \longrightarrow \cr
\psi_{\tilde q \alpha} \;\; (1/2,0,0)^1 \;\; & \longrightarrow \cr
F_i^\dagger \;\; (0,0,1/2)^{-2} \;\; & \longrightarrow \cr}
\quad
\eqalign{
& M^\alpha \;\; (1/2,0)^0, \cr
& \mu_\alpha \;\; (1/2,0)^1, \cr
& v_{\dot\alpha} \;\; (0,1/2)^{-1}, \cr
& K^\alpha \;\; (1/2,0)^{2}, \cr
& \overline M_\alpha \;\; (1/2,0)^0, \cr
& \overline v_{\dot\alpha} \;\; (0,1/2)^{-1}, \cr
& \overline \mu_\alpha \;\; (1/2,0)^1, \cr
& \overline K_\alpha \;\; (1/2,0)^{-2}. \cr}
\eqn\paloma
$$

The $Q$ transformations of the twisted fields are obtained
in the same way as in the case of the $N=2$ vector multiplet.
The resulting transformations, however, are not the ones in
\pera.  First of all notice that the auxiliary fields of the
twisted theory in \paloma\ are different than the auxiliary
fields in \pera. This is a first hint on the existence of
some differences
between the theory in \curva\ and the twisted theory.
Auxiliary fields are useful in supersymmetry because they
permit to close the supersymmey off-shell. In this section
we have considered a version of the $N=2$ hypermultiplet
which contains a minimal set of auxiliary fields.
For this version, however, there is a non-trivial central charge $Z$.
This is an inconvenient for the twisting because then
one finds $Q^2=Z$ instead of $Q^2=0$. On the other hand,
if one disregards this problem and
goes along considering the twisted theory, it turns
out that after integrating the auxiliary fields the resulting action
of the twisted theory is just the action $S$ in \curva\ for a
specific value of $a$:
$$
a={1\over \sqrt{2}}.
\eqn\laa
$$
This can be obtained comparing \curva\ to the action
originated from the twisting of the $N=2$ theory
presented in [\alabas].
This equivalence proves that in the twisted
theory the auxiliary content of the theory and the $Q$-transformations
involving these fields can  be changed
in such a way that an off-shell action can be written
as a $Q$-exact quantity and, furthermore, $Q^2=0$.
In other words, one can indeed affirm that the twisted
theory is topological. This was observed for the first time in [\rok].

Let us briefly describe the $N=2$ hypermultiplet from
the point of view of $N=1$ superspace. This multiplet
contains two $N=1$ chiral multiplets and therefore it
can be described by two $N=1$ chiral superfields
$Q$ (this $Q$ should not be confused with the BRST operator)
and $\widetilde Q$, \ie, these superfields satisfy the
constraints $\overline D_{\dot\alpha} Q=0$ and
$\overline D_{\dot\alpha} \widetilde Q=0$. They have
$U(1)_{\cal R}$ charge $0$. While the superfield $Q$ is in the
fundamental representation of the gauge group, the
superfield $\widetilde Q$ is in the corresponding conjugate
representation. The component fields of these
$N=1$ superfields are:
$$
\eqalign{
Q, \;\; Q^\dagger
 \;\; & \longrightarrow \;\;
q^1, \;\;   \psi_{q\alpha}  \;\;  F^1, \;\; q_1^\dagger \;\;
\overline\psi_{\tilde q \dot\alpha}, \;\;  F_1^\dagger, \cr
\widetilde Q , \;\; \widetilde Q^\dagger
\;\;& \longrightarrow \;\;  q_2^\dagger, \;\; \psi_{\tilde q \alpha},
\;\; F_2^\dagger, \;\; q^2, \;\;
\overline \psi_{q\dot\alpha},   \;\;  F^2.\cr}
\eqn\compas
$$
Again, notice that the $U(1)_{\cal R}$ transformations of the
$N=1$ superfields,
$$
Q \rightarrow Q(\ex^{i\phi}),\;\;\; {\rm and} \;\;\;
\widetilde Q \rightarrow \widetilde Q(\ex^{i\phi}),
\eqn\pertiga
$$
are consistent with the assignment in \paloma.

In $N=1$ superspace the action for the $N=2$ hypermultiplet
coupled to $N=2$ supersymmetric Yang-Mills takes the form:
$$
\int d^4x\,d^2\theta\,d^2\overline\theta\,
(Q^\dagger \ex^V Q + \widetilde Q^\dagger \ex^{-V} \widetilde Q)
+\sqrt{2} \int  d^4x\,d^2\theta\, \widetilde Q \Phi Q +
\sqrt{2}\int d^4x\,d^2\overline\theta\,
\widetilde Q^\dagger \Phi^\dagger Q^\dagger.
\eqn\silvana
$$
Notice that the last two
terms are consistent with the fact that while $\Phi$ is in the
adjoint representation of the gauge group, the superfields
$Q$ and $\widetilde Q$ are in the fundamental and in its
conjugate, respectively.

\section{Twist on K\"ahler manifolds}

In this subsection we describe some aspects of the theory
under consideration when the manifold $X$ is K\" ahler.
Work on Donaldson-Witten theory on K\" ahler manifolds
can be found in
\REF\parkdos{J.S. Park\journal\cmp&163(94)113\journal\np&B423(94)559}
[\wjmp,\parkdos].
When the metric on the four-manifold $X$ is K\"ahler the global
group $SU(2)_L\otimes SU(2)_R$ becomes $U(1)_L\otimes SU(2)_R$
being $U(1)_L$ a subgroup of $SU(2)_L$. The two dimensional
representation of $SU(2)_L$ decomposes under $U(1)_L$ as a sum
of one dimensional representations.
This means that the components $M^1$ and $M^2$ transform
in definite representations of $U(1)_L$ with opposite
charges. In other words, $S^+\otimes E$ has a decomposition
into $(K^{1\over 2}\otimes E) \oplus (K^{-{1\over 2}}\otimes E)$,
where $K$ is the canonical bundle.
The complex structure on $X$ allows to have
well defined complex forms of type $(p,q)$.
We define this complex structure stating the following assignment:
$$
\eqalign{
(\sigma_m)_{1 \dot\alpha}\, d x^m,
\;\;\;\;\; & {\rm type} \;\; (0,1),\cr
(\sigma_m)_{2 \dot\alpha}\, d x^m,
\;\;\;\;\; & {\rm type} \;\; (1,0).\cr}
\eqn\complex
$$
This implies that $(\sigma_{mn})_{\alpha\beta}\,dx^m\wedge dx^n$
can be regarded as a $(0,2)$ form when $\alpha=\beta=1$,
as $(2,0)$ form when $\alpha=\beta=2$, and as a $(1,1)$
form when $\alpha=1,\;\beta=2$.

Let us recall that in the process of twisting the BRST operator
$Q$ was obtained form the supersymmetric charge $Q^i_\alpha$
after identifying $Q^i_\alpha \longrightarrow Q_\alpha{}^\beta$
and then performing the sum $Q=Q_1{}^1+Q_2{}^2$. In the K\"ahler case,
each of the components, $Q_1{}^1$ and $Q_2{}^2$, transforms under
definite $U(1)_L$ representations and therefore one can define
two BRST charges $Q_1=Q_1{}^1$ and $Q_2=Q_2{}^2$. Of course, from
the supersymmetry algebra follows that $Q_1^2=0$ and $Q_2^2=0$.
Furthermore, from their construction: $Q=Q_1+Q_2$.
The action of each of these two operators on the fields is easily
obtained from the supersymmetry transformations. One just
have to set $\eta^{\dot\alpha i}=0$ and, for $Q_1$
$\eta_1^\alpha=\rho_1 \delta_1^\alpha$
and $\eta_2^\alpha=0$, while, for $Q_2$,
$\eta_1^\alpha=0$ and $\eta_2^\alpha=\rho_2\delta_2^\alpha$.
{}From the point of view of $N=2$ superspace the operators $Q_1$ and $Q_2$
can be regarded as a specific derivative respect to some of the $\theta$'s.
In the formulation of the theory on $N=1$ superspace the operator
$Q_1$ can be identified as the derivative respect to $\theta_1$.
This observation will be very helpfull in proving the invariance
under $Q_1$ of the twisted theories.

On a K\"ahler manifold each of the fields on the right hand side of
\twist\ splits into fields which can be thought as components of forms
of type $(p,q)$.
%One finds:
%$$
%\eqalign{
%A_{\alpha\dot\alpha} \;\; & \longrightarrow  \cr
%\lambda_\alpha^i \;\;&\longrightarrow  \cr
%\overline\lambda_{\dot\alpha i} \;\;& \longrightarrow  \cr
%B \;\;& \longrightarrow \cr
%B^\dagger \;\;&\longrightarrow \cr
%D_{ij} \;\;& \longrightarrow \cr}
%\quad
%\eqalign{
% A_{\alpha\dot\alpha} \;\;  & \longrightarrow\cr
% \eta,  \; \chi_{\alpha\beta} \;\;    & \longrightarrow\cr
% \psi_{\alpha\dot\alpha} \;\; & \longrightarrow \cr
% \lambda \;\;  & \longrightarrow\cr
% \phi \;\;  & \longrightarrow\cr
% H_{\alpha\beta} \;\; & \longrightarrow\cr}
%\quad
%\eqalign{
%& A_{1,\dot\alpha} \;\; (1,0), \;\;\; A_{2\dot\alpha} \;\; (0,1), \cr
%& \eta \;\; (0,0), \;\;\; \chi_{11} \;\; (2,0),
%\;\;\; \chi_{12} \;\; (1,1), \;\;\; \chi_{22} \;\; (0,2), \cr
%& \psi_{1,\dot\alpha} \;\; (1,0), \;\;\; \psi_{2\dot\alpha} \;\; (0,1), \cr
%& \lambda \;\; (0,0),\cr
%& \phi \;\; (0,0),\cr
%& H_{11} \;\; (2,0), \;\;\; H_{12} (1,1), \;\;\; H_{22} \;\; (0,2), \cr}
%\eqn\twistka
%$$
%where the type of form appearing for each field has been indicated.
For the matter fields on the right hand side of \paloma\ one just
has the standard decomposition of $S^+\otimes E$ into
$(K^{1\over 2}\otimes E) \oplus (K^{-{1\over 2}}\otimes E)$. For
example
for the field $M^\alpha$ one has:
$$
\eqalign{
M^\alpha & \rightarrow M^1 \in \Gamma(K^{1\over 2}\otimes E),\;
\;\; M^2 \in \Gamma(K^{-{1\over 2}}\otimes E), \cr
\overline M_\alpha & \rightarrow \overline M_1 \in
\Gamma(K^{-{1\over 2}}\otimes \overline E), \;\;\;
\overline M_2 \in
\Gamma(K^{{1\over 2}}\otimes \overline E).\cr}
\eqn\rosi
$$
A similar decomposition holds  for the rest of the fields
in $\Gamma(S^+\otimes E)$ on the right hand side of \paloma.
Notice that the product of an element of
$\Gamma(K^{1\over 2}\otimes E)$ times an element of
$\Gamma(K^{{1\over 2}}\otimes \overline E)$ is a gauge
invariant form of type $(2,0)$. From the identifications in
\paloma\ and \compas\ follows that
the first component of $\widetilde Q Q$, \ie,
$\widetilde Q  Q| = q_2^\dagger  q^1 =
\overline {M_2}^\dagger  M^1$ is a $(2,0)$ form.
Therefore, superpotentials of the form
$\widetilde Q Q$, or $\widetilde Q \Phi Q$ as the one in \silvana\
can be regarded as $(2,0)$-forms.
This is consistent with the observation made
in [\wjmp, \wv] that superpotential terms of a twisted theory on a K\"ahler
manifold must transform as $(2,0)$-forms.

Since the twisted theory obtained fron \superespacio\ and \silvana\
and the topological theory \curva\ are equivalent
on-shell for $a=1/\sqrt{2}$
we will work out the on-shell $Q_1$-transformations for this case.
Notice that only if $a=1/\sqrt{2}$ in \curva\ one can guarantee
$Q_1$-invariance.
The $Q_1$-transformations for the twisted fields in the
$N=2$ vector multiplet turn out to be:
$$
\eqalign{
&[Q_1,A_{1\dot\alpha}]=\psi_{1\dot\alpha},\cr
&[Q_1,A_{2\dot\alpha}]=0,\cr
&\{Q_1,\psi_{1\dot\alpha}\}=0,\cr
&\{Q_1,\psi_{2\dot\alpha}\}=D_{2\dot\alpha}\phi,\cr
&[Q_1, \phi]=0,\cr }
\quad
\eqalign{
&[Q_1, \lambda]={1\over 2}\eta+\chi_{12},\cr
&\{Q_1, {1\over 2}\eta+\chi_{12}\}= 0,\cr
&\{Q_1, {1\over 2}\eta^a-\chi_{12}^a\}=
+2i(F^{a+}_{12}+{i\over 2}(\overline M_1 T^a M_2
+\overline M_2 T^a M_1))-{i\over 2}[\lambda,\phi]^a,\cr
&\{Q_1,\chi_{11}^a \}=\overline M_1 T^a M_1 , \cr
&\{Q_1,\chi_{22} \}= -iF^{+}_{22} , \cr}
\eqn\sandia
$$
where we have used that the generators of the gauge group
are normalized in such a way that $\tr(T^aT^b)=\delta^{ab}$.
For the matter fields one finds:
$$
\eqalign{
& [Q_1, M^1]   = \mu^1, \cr
& [Q_1, M^2]   = 0, \cr
& \{ Q_1, \mu^1 \}   = 0,      \cr
& \{ Q_1, \mu^2 \}   = \phi M^2, \cr
& \{ Q_1, v_{\dot\alpha}\}   = D_{2\dot\alpha}M^2,\cr}
\qquad\qquad
\eqalign{
& [Q_1, \overline M_1]   = 0 , \cr
& [Q_1, \overline M_2]   = \overline \mu_2, \cr
& \{ Q_1, \overline \mu_1 \}   = \phi \overline M_1, \cr
& \{ Q_1, \overline \mu_2 \}   = 0 ,      \cr
& \{ Q_1, \overline v_{\dot\alpha}\}   = \overline D_{2\dot\alpha}
\overline M_1.\cr}
\eqn\elipse
$$
It is straightforward to verify that indeed $Q_1^2=0$
on-shell after working out the transformations
of the different components of $F_{\alpha\beta}^+$:
$$
[Q_1, F_{11}^+] =i D_{1\dot\sigma}
\psi_1{}^{\dot\sigma}, \;\;\; [Q_1, F_{12}^+]={i\over 2}
D_{2\dot\sigma}
\psi_1{}^{\dot\sigma}, \;\;\;
[Q_1, F_{22}^+]=0.
\eqn\melon
$$

The $Q_2$-transformations are easily computed from
\sandia, \elipse\ and \pera\ after using $Q=Q_1+Q_2$.
The action $S$ in \curva\ for $a=1/\sqrt{2}$ is invariant under
both, $Q_1$ and $Q_2$ symmetries. This can be verified
explicitly or just using the following argument
based on $N=1$ superspace.
On the one hand, the topological action \curva\ can be regarded as
a twisted version of the sum of the $N=1$ superspace actions
\superespacio\ and \silvana. On the other hand, the $Q_1$ operator
is equivalent to a $\theta_1$-derivative. Acting with
this derivative on \superespacio\ and \silvana\ one gets zero: for the
terms involving chiral fields one ends with two many
$\theta$-derivatives, while for the other terms one just gets
a total derivative after using the fact that
$[D_\alpha,\overline D^2] = i \partial_{\alpha\dot\alpha}
\overline D^{\dot\alpha}$.

It is often convenient to regard the
the observables $I(\Sigma)$ in  \moreobser\
in terms of the Poincar\'e dual of the homology
cycle $\Sigma$:
$$
I(\Sigma) = \int_\Sigma {\cal O}^{(k)} =
\int_X {\cal O}^{(2)}\wedge [\Sigma],
\eqn\leon
$$
where $[\Sigma]$ denotes the Poincar\'e dual. On K\" ahler manifolds,
$I(\Sigma)$ can be decomposed
in three different types of operators depending on
which holomorphic part of ${\cal O}^{(2)}$ is taken into account.
If only the $(p,q)$ part ($p+q=2$) of ${\cal O}^{(2)}$ is considered
we will denote the corresponding operator by
$I^{p,q}(\Sigma)$. For example, for the $(1,1)$ part:
$$
I^{1,1}(\Sigma) = {1\over 2}\int_X e (i \phi F_{12} -
{1\over 2} \psi_{1\dot\alpha}
\psi_2^{\dot\alpha}) [\Sigma]_{12} =
{1\over 2}\int_X e (i B^{\dagger} F_{12} -
{1\over 2}\overline \lambda_{1\dot\alpha}
\overline \lambda_2^{\dot\alpha}) [\Sigma]_{12},
\eqn\launouno
$$
where in the last step we have used \twist, and we have denoted
by $\Sigma_{12}$ the $(1,1)$ part of $\Sigma$.

\section{The perturbed massive theory on K\"ahler manifolds}

One of the main ingredients in the analysis made by
Witten in [\wjmp] is the existence of a perturbation
of the twisted $N=2$ Yang-Mills theory on K\"ahler manifolds
which  while preserving the topological character
of the twisted theory it  allows to regard the theory
from an untwisted point of view as an $N=1$ supersymmetric theory.
Witten achieved this demostrating that on a K\"ahler
manifold it is possible to add an $N=1$ supersymmetric
 mass-like term for the chiral superfield $\Phi$ while
keeping the topological character of the theory.
In this subsection we will show that this is also possible
for the topological quantum field theory
which describes non-abelian monopoles. Notice that, as we need the
superpotentials to transform as $(2,0)$-forms, to generate a mass term for
$\Phi$ we must pick a holomorphic $(2,0)$-form on $X$. This is not always
possible on an arbitrary K\"ahler manifold,
but as we are assuming $b_2^+ >1$
in order to have a well-defined moduli problem,
we guarantee that $H^{2,0}(X)
\not= 0$ and hence that such a form exists.

Let us consider a holomorphic $(2,0)$ form $ \omega$ on $X$.
Its only non-vanishing component is:
$$
\omega_{11} = (\sigma_{lk})_{11} \omega_{mn} \epsilon^{lkmn}.
\eqn\forma
$$
We will denote the unique non-vanishing component of the
$(0,2)$ form $\overline\omega$, conjugate to $\omega$,
by $\overline\omega_{22}$ ( $\overline\omega_{22}=
(\omega_{11})^*$)

Following [\wjmp] we begin making a perturbation of
the action $S$ in \action\ by adding a term of the form,
$$
I(\omega) = \int_X {\cal O}^{(2)} \wedge
\omega,\eqn\primer
$$
where $I(\omega)$ is the observable defined in \moreobser.
In \primer\ we are denoting the Poincar\'e dual to the holomorphic
$(2,0)$ form $\omega$ by the symbol $\omega$ as well.
Using the $Q_1$-transformations \sandia\ and \elipse,
this term can be written as:
$$
I(\omega)= -{1\over 2}\int_X d^4 x \, e \,
\omega_{11} \tr({1\over 2}\psi_{2\dot\alpha}\psi_2^{\dot\alpha}) +
\big\{ Q_1, -{1\over 2}  \int d^4 x \, e \,
\omega_{11} \tr(\phi\chi_{22}) \big\}.
\eqn\tomate
$$
The first part of this term indicates some progress
towards the construction of an $N=1$ mass-like term.
However, \tomate\ is not invariant under $Q_1$.
Contrary to the case of the theory without matter fields
this term is not even $Q_1$-invariant on-shell. One can
remedy this problem if instead of introducing $I(\omega)$
one considers:
$$
\tilde I(\omega) = I(\omega) -{1\over 2}  \int_X d^4 x \, e \,  \omega_{11}
\overline M_2 \phi M_2.
\eqn\flordos
$$
Indeed, $\{ Q_1, \tilde I(\omega) \}$ turns out to be
proportional to the field equation resulting after
making a variation respect to $\chi^{22}$ in the twisted action
$S$ in \curva\ (with $a=1/\sqrt{2}$) .

The term $\tilde I(\omega)$ implies further progress
towards the perturbation by an $N=1$ supersymmetric
mass term. Notice that from an $N=1$ superspace point of
view we intend to obtain a term of the form,
$$
m \int d^4x \, d^2 \theta \tr(\Phi^2) + \overline m \int d^4x \, d^2
\overline  \theta \tr({\Phi^\dagger}^2) .
\eqn\masa
$$
This type of term, added to a theory which already has the last
two terms in \silvana, leads, when written in component fields, to
terms like the one added to $I(\omega)$ in \flordos.

To make further progress in the perturbation towards an $N=1$
supersymmetric mass
term while maintaining the $Q_1$ symmetry we will modify
the $Q_1$-transformation of $\chi_{11}$ in the following way:
$$
\{Q_1, \chi_{11}^a \} \longrightarrow
\{Q_1', \chi_{11}^a \} = \overline M_1
T^a M_1+ \, \omega_{11} \phi^a,
\eqn\tinte
$$
while for the rest of the fields the action of $Q_1$ and
$Q_1'$ remains the same. Notice that still one has
${Q_1'}^2=0$ on-shell.

Under $Q_1'$ the action $S$ is not invariant. However,
one can verify that now the perturbed action
$S+\tilde I(\omega)$ is invariant and not a field equation
as before. On the other hand,
adding a $Q_1'$-exact term will keep the $Q_1'$-invariance
of the theory. It is rather remarkable that adding just the term,
$$
-{1\over 2}\Big\{Q_1', \int d^4 x\,e \, \overline
\omega_{22} \tr(\lambda \chi_{11}) \Big\},
\eqn\lugo
$$
one finds that the perturbed action is just the action $S$ plus an
$N=1$ supersymmetric mass term for the chiral superfield $\Phi$:
$$
\eqalign{
S+&\tilde I(\omega) + \{ Q_1',...\}\cr
&=S-{1\over 2}\int_X d^4 x\, e \Big(\overline \omega_{22}
\tr\big(({1\over 2}\eta+\chi_{12})\chi_{11}\big)+
\omega_{11} \tr( {1\over 2}\psi_{2\dot\alpha} \psi_2{}^{\dot\alpha}))
+  \int_X d^4 x\, e\, \omega_{11}\overline\omega_{22} \tr(\lambda \phi)\cr
&\;\;\;\;\;\;\;\;\;-{1\over 2} \int_X d^4 x \, e
(\omega_{11} \overline M_2 \phi M_2 +
\overline\omega_{22} \overline M_1 \lambda M_1).\cr}
\eqn\marmol
$$
This perturbation of the action $S$  contains all the terms
present in the $N=1$ supersymmetric mass term \masa\
after setting $m=\omega_{11}$ and integrating out the auxiliary
fields. Writing the twisted fields in
terms of the untwisted ones the form of \masa\ in component
fields is obtained. Recall that according to \twist\ and \debbie,
$\lambda_2^2={1\over 2}\eta+\chi_{12}$, $\lambda^2_1=\chi_{11}$,
$\lambda_{2\dot\alpha}=\psi_{2\dot\alpha}$, $B^\dagger=\phi$,
and $B=\lambda$. For the matter fields one can read their
untwisted counterparts from \paloma.

Our analysis implies that if one denotes correlation
functions of observables in the twisted theory by
$\langle A_1\cdots A_n\rangle$, and in the perturbed
theory by  $\langle
A_1\cdots A_n\rangle_1$, the relation between them is:
$$
\langle A_1\cdots A_n\rangle_1
=\langle A_1\cdots A_n \ex^{-\tilde I(\omega)} \rangle.
\eqn\lawi
$$
As argued in [\wjmp], given some  homology cycles $\Sigma$,
it can be assumed that near their intersection
they look like holomorphically embedded Riemann surfaces.
This means that actually the only relevant part of the
two-form operators entering \lawi\ are of type $(1,1)$.
Precisely those are the two-form operators invariant under $Q_1'$.
This follows trivially using \sandia\ in \launouno.
As the zero-form observables in \lawi\ are also
invariant under $Q_1'$ one can regard the right hand
side of \lawi\ as a topological quantum field theory
whose BRST operator is $Q_1'$ and its action
is $S+\tilde I(\omega)$.

The effect of
an extra term $I(\omega)$ in the action
of Donaldson-Witten theory was
studied by Witten in [\wjmp]. He showed that its effect
on correlators of  observables can be described as a
shift on the parameters corresponding
to the observables containing two-form operators.
We will finish this section
showing that the relevant contribution from $\tilde I(\omega)$
in \lawi\ and from $I(\omega)$ in the case of Donaldson-Witten
theory is the same. Therefore, in our theory the effect of
the presence of $\tilde I(\omega)$ in \lawi\ is also a shift
in those parameters.

The quantity $\tilde I(\omega)$ can be written as,
$$
\eqalign{
\tilde I(\omega) =&{1\over 2}
\int_X d^4 x\, e\, \omega_{11} \Big(i\phi^a(F^{a+}_{22}+i\overline M_2
T^a M_2)-\tr({1\over 2}\psi_{2\dot\alpha}\psi_2{}^{\dot\alpha})\Big)\cr
=&-{1\over 2}
\int_X d^4 x\, e\, \omega_{11} \Big(\tr(
\{Q,\phi\chi_{22}\}
+{1\over 2}\psi_{2\dot\alpha}\psi_2{}^{\dot\alpha})\Big),\cr}
\eqn\loro
$$
after using \pera. This means that the vacuum expectation values
on the right hand side of \lawi\ can be written as
$$
\langle A_1\cdots A_n \ex^{J(\omega)} \rangle.
\eqn\lawidos
$$
where,
$$
J(\omega) =  {1\over 4}
\int_X d^4 x\, e\, \omega_{11}
\tr(\psi_{2\dot\alpha}\psi_2{}^{\dot\alpha}).
\eqn\francesca
$$
This is precisely the same expression that one obtains in
the case of Donaldson-Witten theory. Notice that in that case
$F_{22}^+$ is $Q$-exact and one has the same $Q$-transformations
as in our theory for the field $\psi_{a\dot\alpha}$.

Another argument to show that the presence of the term involving
the massive fields in $\tilde I (\omega)$ is irrelevant is just
to point out that the contributions from the functional integral
on the right hand side of \lawi\ are localized  on  configurations
satisfying the monopole equations. As  shown in [\nonab]
the $(0,2)$ part of those equations implies $\overline M_2 T M_2=0$.

\endpage
\chapter{Vacuum structure of the $N=2$ and $N=1$ theories}

As we have seen, when we perturb
the original $N=2$ theory
to a $N=1$ theory on a K\"ahler manifold, the resulting
theory preserves the topological symmetry and one can compute the
topological correlation functions in the $N=1$ theory.
If this theory has a mass
gap and presents topological invariance, the only relevant
information we need in this computation is the
structure of the $N=1$ vacua  and their
symmetries, as was shown by Witten in
[\wjmp]. When we consider $N=2$ pure
Yang-Mills theory perturbed by a mass term
for the chiral multiplet
$\Phi$, we know that at low energies we are reduced to a
$N=1$ Yang-Mills theory. This theory is supposed to have a mass gap and its
vacua have the symmetry pattern coming from spontaneous chiral symmetry
breaking [\witsusy]. But the standard
conjectures about the structure of vacua
of this theory can be obtained from the structure of the quantum
moduli space of vacua of the
corresponding $N=2$ theory, as it has been shown in
[\wspure]. The same method can be applied to the $N=2$ theory with matter
perturbed by the $N=1$ mass term for $\Phi$ [\wsmatter].
In this section we will
use the information about symmetries
and vacua of the $N=2$ theory [\wsmatter]
to show that in fact the $N=1$ theory we are dealing with has a mass gap
and we will obtain a precise description of its vacua.

$N=2$ supersymmetric QCD with
gauge group $SU(N_{c})$ and $N_{f}$ hypermultiplets in the fundamental
representation of the
gauge group has the  $U(1)_{\cal R}$ symmetry
described in \salto\ and \pertiga:
$$
\eqalign{
&W_{\alpha} \longrightarrow
{\rm e}^{-i\phi}W_{\alpha}({\rm e}^{i\phi} \theta ),
\cr &\Phi \longrightarrow
{\rm e}^{-2i\phi} \Phi ({\rm e}^{i\phi}\theta),\cr}
\qquad\qquad
\eqalign{
& Q^{i}  \longrightarrow Q^{i}({\rm e}^{i\phi}\theta),\cr
& {\widetilde Q}_{\tilde i} \longrightarrow {\widetilde Q}_{\tilde i} ({\rm
e}^{i\phi}\theta).\cr}
\eqn\lola
$$
In component fields the corresponding transformations can be
read from \twist\ and \paloma:
$$
\eqalign{
&\lambda^{1},\; \lambda^{2}
\longrightarrow {\rm e}^{-i\phi} \lambda^{1},\;
{\rm e}^{-i\phi}  \lambda^{2}, \cr
&B \longrightarrow {\rm e}^{-2i\phi} B,\cr
&\psi_{qi},\; \psi_{\tilde q \tilde i} \longrightarrow
{\rm e}^{i\phi} \psi_{qi},\;
{\rm e}^{i\phi} \psi_{\tilde q \tilde i}.\cr}
\eqn\maslola$$
 This symmetry is anomalous because of instanton effects. The anomaly is
$4N_{c}-2N_{f}$ ($2N_{c}$ from $\lambda^1$
and $\lambda^2$, which live in the adjoint
representation of the gauge group,
and $2$ from each couple of fermions $\psi_{q}$,
$\psi_{\tilde q}$ in the hypermultiplet). In the case we are dealing with,
namely $N_{c}=2$ and $N_{f}=1$
(which gives the $SU(2)$ monopole equations) the anomaly
is $6$ and we should expect the
${\bf Z}_{6}$ anomaly-free discrete subgroup:
$$
\eqalign{
&\lambda^{1},\; \lambda^{2} \longrightarrow
{\rm e}^{-{i\pi \over 3}}
\lambda^{1}, \; {\rm e}^{-{i\pi \over 3}} \lambda^{2}, \cr
&B \longrightarrow {\rm e}^{-{2i\pi \over 3}}
B,\cr
&\psi_{q},\; \psi_{\tilde q}
\longrightarrow  {\rm e}^{i\pi\over 3} \psi_{q},
\;{\rm e}^{i\pi\over 3} \psi_{\tilde q}.\cr}
\eqn\enedos
$$
However, since we are considering one hypermultiplet in the
fundamental representation of $SU(2)$, we must take into account
that the quark $Q$
and the antiquark ${\widetilde Q}$
live in isomorphic representations of the gauge group
(for $SU(2)$, ${\bf 2}
\simeq {\bf {\bar 2}}$). As a
consequence of this isomorphism, and when the matter fields are massless,
we have a parity symmetry $\rho$ interchanging the quark and the antiquark:
$$
\rho : Q \leftrightarrow {\widetilde Q}.
\eqn\parida$$
This symmetry is anomalous, as can be
seen from  the 't Hooft interaction term,
$$
(\lambda^1)^4 (\lambda^2)^4 \psi_{q} \psi_{\tilde q}.
\eqn\tof$$
Nevertheless, one
can combine the $\rho$ symmetry with the square
root of ${\bf Z}_{6}$ in \enedos\ to obtain
an anomaly-free ${\bf Z}_{12}$ subgroup.

Under the ${\bf Z}_{12}$ symmetry the quantity $u={\rm Tr}
B^2$ transforms as $u \rightarrow {\rm e}^{-2\pi i /3}u$
and gives a global ${\bf
Z}_{3}$ symmetry on the $u$ plane.
This plane parametrizes in fact the moduli space of
vacua of the theory.
Classically, $SU(2)$ is broken to $U(1)$ for $u \not= 0$, and at
$u=0$ the gauge symmetry is unbroken
as the gluons become massless. Quantum mechanically,
the picture that emerges is very different:
there are three singularities interchanged by the ${\bf
Z}_{3}$ symmetry.
These singularities are points where magnetic monopoles or dyons
become massless, and the description
based on a low-energy effective action which
includes only the photon multiplet of the unbroken $U(1)$ breaks down: an
additional massless hypermultiplet
must be included near each singularity [\wsmatter]. The
effective theory becomes therefore
$N=2$ supersymmetric QED with a massless hypermultiplet.

The superpotential of the resulting
$N=2$ supersymmetric QED in terms of $N=1$ superfields has the following
form:
$$
W_{\rm M}={\sqrt 2}A M {\widetilde M} .
\eqn\superab
$$
In this expression $A$ denotes the $N=1$ chiral
multiplet of the $N=2$ Yang-Mills
field (it is the abelian analogue of $\Phi$). The superfields $M$ and
${\widetilde M}$ are the $N=1$ chiral multiplets which represent the $N=2$
hypermultiplet in the abelian case. They have opposite charges. We are
interested in the vacuum structure of
the $N=1$ theory which is obtained when one
adds a mass term $m{\rm Tr} \Phi^2$
to the $N=2$ theory with matter. For this one can
use the effective low-energy action,
as it is shown in [\wspure, \wsmatter]. The
effective contribution of the mass
term for $\Phi$ in the low-energy theory can be
represented by an additional
term in the superpotential $W_{\rm eff}=mU$, where $U$
is a chiral superfield whose
first component is the operator $u$. The vacua of the
$N=1$ theory are given by the
critical points of the superpotential, up to
complexified $U(1)$ gauge
transformations (this is equivalent to set the $D$ terms to
zero and divide by $U(1)$). At non-singular points of
the moduli space, $W=W_{\rm eff}$ and therefore,
if one supposes that $du \not= 0$, there are no
supersymmetric ground states at all. The only points in the quantum moduli
space of vacua of the $N=2$ theory which give rise
to $N=1$ vacua are precisely the
singularities where monopoles
become massless. In this case $W=W_{\rm
M}+W_{\rm eff}$ and there are
critical points where magnetic monopoles get an
expectation value.
Hence, the resulting $N=1$ theory has three vacua related by the
${\bf Z}_{3}$ symmetry of the $u$-plane.
In these points one can also check that
there is mass gap and condensation of monopoles.

Because of the mass gap of the
$N=1$ theory  we can use the
physical properties of this kind of theories to
evaluate the correlation functions.
We also know that this theory has three vacua,
but to have a clear picture of their symmetries we need the resulting
$U(1)_{\cal R}$
symmetry of the perturbed theory.
Notice that the mass term for the $\Phi$ field
breaks the second transformation in \lola\ due to the presence
of the fermionic fields
$\lambda^2$, as can be seen from
\marmol. Thus under the new $U(1)_{\cal R}$ symmetry we
must have:
$$
\Phi  \longrightarrow {\rm e}^{-i\phi} \Phi ({\rm
e}^{i\phi}\theta),
\eqn\orense
$$
and this in turn imposes, because of the superpotential
term, the following
transformation for the matter fields:
$$
\eqalign{ & Q
\longrightarrow {\rm e}^{-i\phi/2}
Q({\rm e}^{i\phi}\theta),\cr & {\widetilde Q}
\longrightarrow {\rm e}^{-i\phi /2}{\widetilde Q}
({\rm e}^{i\phi}\theta).\cr}
\eqn\uonemat
$$
Rescaling the charges to make them integers, we have the
following $U(1)_{\cal R}$ symmetry for
the perturbed theory in terms of components
fields:
$$
\eqalign{ & \lambda^{1},\; B \longrightarrow {\rm e}^{-2i\phi}
 \lambda^{1},\; {\rm e}^{-2i\phi}  B,
\cr &
q, \; {\tilde q} \longrightarrow {\rm e}^{-i\phi} q,
\; {\rm e}^{-i\phi} {\tilde q},
 \cr
&\psi_{q}, \; \psi_{\tilde q} \longrightarrow  {\rm e}^{i\phi} \psi_{q},
\; {\rm e}^{i\phi} \psi_{\tilde q}.\cr}
\eqn\uonefin
$$

The anomaly-free discrete subgroup of the transformations
\uonefin\ is ${\bf Z}_{6}$. However,
one must take into account the $\rho$ symmetry \parida, as the
addition of the mass term for $\Phi$ doesn't break it. Again, we have an
enhancement of the discrete symmetry to ${\bf Z}_{12}$. The resulting
transformations are:
$$
\eqalign{
& \lambda^{1},\; B \longrightarrow {\rm e}^{-\pi i/3} \lambda^1,
\; {\rm e}^{-\pi i/3}  B, \cr
& q \longrightarrow {\rm e}^{-\pi i/6}{\tilde q} ,\,\,\,\,\ {\tilde
q} \longrightarrow {\rm e}^{-\pi i/6}q,\cr
 &\psi_{q} \longrightarrow  {\rm e}^{\pi i/6} \psi_{\tilde q},\,\,\,\,\
 \psi_{\tilde q} \longrightarrow  {\rm e}^{\pi i/6} \psi_{q}.\cr}
\eqn\uonedef
$$
These transformations leave invariant the 't Hooft
term $(\lambda^1)^4 \psi_{q} \psi_{\tilde
q}$. We know that this theory has
only three vacua, and therefore there must
be spontaneous symmetry breaking of \uonedef, as it happens
in the pure $N=1$ Yang-Mills theory.
To identify the pattern of this breaking, notice
that these vacua are labeled by the order parameter $u={\rm Tr}
B^2$. It is easy to see that
the unbroken symmetry which gives this vacuum structure is:
$$\eqalign{
&\lambda^1 \longrightarrow -\lambda^1,\cr
& \psi_{q} \longrightarrow  i \psi_{\tilde q},\cr
&\psi_{\tilde q}\longrightarrow  i \psi_{q}.\cr}
\eqn\martin
$$
This is precisely the maximal  subgroup of \uonedef\ which allows fermion
masses for $\lambda^1$ and for $\psi_{q}$, $\psi_{\tilde q}$ (notice that
the mass term for the matter fields
changes its sign under the parity symmetry involved in  \martin). The
spontaneous chiral symmetry breaking  in \martin\ is induced by a
\vev\ of the
gauge invariant order parameter  $X=\widetilde {Q} Q $ as in
\REF\susybreak{I. Affleck,
M. Dine and N. Seiberg \journal\np&B241(84)493}
[\susybreak].

\endpage
\chapter{Computation of the polynomial invariants}

In this section we compute the topological invariants
corresponding to $SU(2)$ monopoles by two different
methods. The first one is based on the abstract approach
developed in [\wjmp] and is valid only for K\" ahler manifods.
The second method, which is valid for arbitrary
spin manifolds, uses electric-magnetic duality [\wspure,\wsmatter]
and is inspired by the approach developed in [\mfm].

\section{K\"ahler manifolds}
Now that we have the information
about the vacuum structure of the $N=1$ theory and
their symmetries we can compute the correlation functions of the
topological theory on a K\"ahler,
spin manifold. Because of the presence of the mass
gap most of the arguments of
[\wjmp] go through. Although the structure of the mass
perturbation on a K\"ahler manifold
introduces some subtleties which we will consider
later (the cosmic string theory),
in a first approach the correlation functions take
the form:
$$
\langle {\rm exp} (\sum_{a}\alpha_{a}I(\Sigma_{a})+\mu {\cal O})
\rangle=\sum_{\rho}C_{\rho}
{\rm exp} (\gamma_{\rho}v^2 +\mu \langle \rho |{\cal O}|
\rho \rangle). \eqn\corr
$$
In this expression the sum is  over the three vacua $|\rho \rangle$ labeled
by the index $\rho =1,2,3$; $v=\sum_{a}\alpha_{a}[\Sigma_{a}]$, where
$[\Sigma_{a}]$ is the cohomology class Poincar\'e dual to $\Sigma_{a}$, and
$v^2=\sum_{a,b}\alpha_{a}\alpha_{b}\sharp(\Sigma_{a}\cap \Sigma_{b})$, where
$\sharp (\Sigma_{a}\cap \Sigma_{b})$  is the intersection number of
$\Sigma_{a}$ and $\Sigma_{b}$. The constant  $C_{\rho}$ is the partition
function in the $\rho$ vacuum, and the mass gap and topological  character
of the theory imply that it must have the structure:
$$
C_{\rho}={\rm exp}(a_{\rho}\chi + b_{\rho}\sigma) .
\eqn\asturias
$$

The constants which
appear in \corr\ are not independent
because the theory has a ${\bf Z}_{3}$ broken
symmetry which relates
the three vacua and is given by \uonedef.  First let us work out
the
relation between the $C_{\rho}$.
As these constants are given by the partition function
of the theory at different vacua,
and the vacua are related by a non-anomalous
symmetry, one should think
that they are equal. But actually, as we are working now on a
curved four-manifold,
the anomalies have gravitational contributions which were not
taken into account in section 3
(where $X={\bf R}^4$), and the path integral measure
does change. We must take into
account also  the new geometrical content of the fields
after twisting. The field ${\overline \lambda_1}$
is now a $(1,0)$-form, and
$\lambda^{1}$ contains a $(2,0)$-form part,
a $(1,1)$ part and a scalar part. The
operator relating them is:
$$
\partial \oplus p_{+}{\overline \partial} \oplus {
\partial}^{\dagger}:\,\ \Omega^{1,0} \longrightarrow
\Omega^{2,0} \oplus \Omega^{1,1} \oplus \Omega^{0},
\eqn\doloresdos
$$
and its index is given by half the
dimension of ${\cal M}_{\rm ASD}$  in \masd\ (notice that
the complex conjugate of this operator
gives the one for ${\overline \lambda}_2$,
which is a $(0,1)$-form, so both indices
are equal and the sum of them gives the
index of the original ASD complex \asd).
Therefore the anomaly due to the first
transformation in \uonedef\ is:
$$
{\rm e}^{ {\pi i \over 3}\{ 4k-{3 \over 4}(\chi +\sigma)\} }.
\eqn\fase
$$
We must take into account also the
transformation of the matter fermions. After twisting
they are spinors, and we have the correspondence $\psi_{q}\rightarrow
\mu$, $\psi_{\tilde q} \rightarrow {\bar
\mu}$ (see \paloma).
Notice that, due to the $\rho$ symmetry in \parida, there is an
additional contribution to the
anomaly coming from this transformation, as we saw in
\tof. Now we must also compute the gravitational part and obtain the total
anomaly (a similar problem is addressed in section 4.4 of
\REF\sab{E. Witten, ``On S-Duality In Abelian Gauge Theories",
hep-th/9505186}
[\sab]). The path integral measure
for the twisted matter fermions can be written as:
$$
\prod _{I} d\mu_{I} d{\bar \mu}_{I} \prod _{J} dv_{J} d{\bar v}_{J},
\eqn\tinta
$$
where the index $I=1, \cdots, \nu_{+}$
refers to the $\mu^{\alpha}$ zero modes (of
positive chirality) and the index
$J=1, \cdots, \nu_{-}$ to the $v_{\dot\alpha}$ zero
modes (of negative chirality).
Under the transformation in \uonedef, the measure \tinta\
transforms as:
$$
(-1)^{\nu_{+}+\nu_{-}}{\rm e}^{ -{\pi i \over 6}2 (k+ {\sigma \over 4})}
=(-1)^{-k-{\sigma \over 4}}{\rm e}^{ -{\pi i \over 3}
(k+ {\sigma \over 4})},
\eqn\disco
$$
where we have taken into account
that $\nu_{+}-\nu_{-}={\rm index}\,\ D = -k-\sigma
/ 4$, according to \dirac. Putting together
both factors we obtain:
$$
(-1)^{\Delta}{\rm e}^{-{\pi i \over 12}\sigma},
\eqn\anomaly
$$
where $\Delta$ was introduced in
\ladelta\ and we have used that $\sigma \equiv 0
\,\ {\rm mod}\,\ 8$.
Notice that the $k$ dependence has dropped out, because the
symmetry in \uonedef\ is not anomalous under Yang-Mills instantons, and
\anomaly\ contains only the gravitational contribution to the anomaly. The
result, in terms of the constants $C_{\rho}$, is:
$$
C_{2}=(-1)^{\Delta}{\rm e}^{-{\pi i \over 12}\sigma}
C_{1},\,\,\,\,\ C_{3}=
{\rm e}^{-{\pi i \over 6}\sigma}C_{1}.
\eqn\zvacios
$$

We would also like to relate the constants
$\gamma_{\rho}$ and the expectation values
$\langle \rho |{\cal O}| \rho \rangle$ in
\corr\ for the different vacua. This is
easily done taken into account the
transformations of the corresponding observables
under the symmetry \uonedef. As it
is argued in [\wjmp], for the observables
$I(\Sigma)$ on a K\"ahler manifold
one can consider only the $(1,1)$ part. If we call
$\alpha$ the generator of the discrete
symmetry in \uonedef\ in the operator formalism,
after taking into account \launouno, one finds
the following relations:
$$
\eqalign{
& \alpha|\rho \rangle = |\rho +1
\rangle, \,\,\,\,\,\ \rho \equiv 1 \,\ {\rm mod} \,\ 3, \cr&
\alpha {\cal O} \alpha^{-1} =
{\rm e}^{2\pi i \over 3} {\cal O},\,\,\,\,\ \alpha
I^{1,1}(\Sigma)  \alpha^{-1} = {\rm e}^{\pi i \over 3} I^{1,1}(\Sigma),
 \cr}\eqn\obanomalia
$$
which lead to:
$$
\eqalign{
&\gamma_{2}={\rm e}^{-{2\pi i \over 3}}
\gamma_1, \,\,\,\,\ \gamma_{3}={\rm e}^{-{4\pi i
\over 3} }\gamma_1, \cr
&\langle 2 |{\cal O}|2
\rangle={\rm e}^{-{2\pi i \over 3}}\langle 1 |{\cal O}|1 \rangle,
\,\,\,\,\ \langle 3 |{\cal O}|3
\rangle={\rm e}^{-{4\pi i \over 3}}\langle 1 |{\cal O}|1
\rangle. \cr}
\eqn\relaciones$$
With these relations we have determined completely
the bulk structure of the vacua,
which comes from the underlying $N=1$ theory.

One has to take into account however that
 the mass perturbation which gives
this theory was done with a $(2,0)$ holomorphic form $\omega$, and the
mass will vanish when this form does.
In general, $\omega$ vanishes on a divisor $C$
representing the canonical class of $X$.
The simplest case is the one in which $C$ is a
union of disjoint Riemann surfaces
$C_y$ of multiplicity $r_{y}=1$ ({\it i.e.},
$\omega$ has simple zeroes along this components),
and therefore the canonical divisor
of $X$ can be written as,
$$
c_{1}(K)=\sum_{y}[C_y].
\eqn\canonico
$$
As discussed in [\wjmp, \wv], near these surfaces $C_{y}$ we
have an effective two-dimensional
theory (the cosmic string theory) with additional
symmetry breaking. In particular,
along the worldsheets of the strings $C_{y}$ each bulk
vacuum bifurcates according to
a new pattern of symmetry breaking. As in [\wjmp, \wv],
we will assume that each bulk
vacuum gives two vacua along the string, and therefore that
there is spontaneous symmetry breaking of $\lambda^{1} \rightarrow
-\lambda^{1}$. As we will see,
this assumption is the most natural one from several points
of view. First of all,
the contributions from the new vacua cooperate with the bulk
structure in such a way
that the resulting expression has the adequate properties.
Second, with this assumption, the final
expression can be naturally understood as a
consequence of electric-magnetic
duality of the underlying $N=2$ theory. Finally,
as we will describe in sect. 5, where we consider the $N=2$
supersymmetric theory with a massive hypermultiplet,
one can show that, although the bulk structure of vacua of that theory
is different from the one under consideration,
the ``internal" structure
of each vacuum corresponds in fact to a two-fold bifurcation.

 Let us briefly explain, following [\wjmp],
what is the effect of the new vacua in the
computation of the correlation functions.
Each bulk vacuum $|\rho \rangle$ leads to two
vacua of the cosmic string
theory $|\rho +\rangle$, $|\rho - \rangle$, which are related
by the broken symmetry $\alpha^{3}$. The
surfaces $C_{y}$ give new contributions
to the correlators through their intersection
with the surfaces $\Sigma_{a}$.
The observables $I(\Sigma_a)$ will be described by
 $\sharp (\Sigma_a \cap C_{y}) V_{y}$, where $V_{y}$ is the
insertion of a cosmic string operator
$V$ on $C_{y}$ which has the same quantum numbers
of $I^{1,1}(\Sigma_a)$. From \launouno\ and \uonedef\ follows that
it transforms under $\alpha^3$ as:
$$
\alpha^3 V \alpha^{-3}=-V.
\eqn\santander
$$
Now, for a given bulk
vacuum $|\rho \rangle$ we must take into account its bifurcation
along the diferent surfaces
$C_{y}$, and compute the vacuum expectation values of ${\rm exp}
(\sum_{a}\alpha_{a}I(\Sigma_{a}))$. The result is [\wjmp]:
$$
 \prod_{y}\Big( {\rm
exp}(\phi_{y}\langle \rho +|V|\rho + \rangle) + t_{y}{\rm
exp}(-\phi_{y}\langle \rho +|V|\rho + \rangle) \Big).
\eqn\cosmicomicas
$$
In this expression,
$\phi_{y}=\sum_{a} \alpha_{a}\sharp (\Sigma_a \cap C_{y})$ and the
factor $t_{y}$ is similar
to \anomaly\ and comes from an anomaly in the two-dimensional
effective theory. It is given by:
$$
t_{y}=(-1)^{\epsilon_{y}},
\eqn\late$$
where ${\epsilon_{y}}$ is $0$
($1$) if the spin bundle of $C_{y}$ is even (odd). The
$\epsilon_{y}$ verify [\wjmp]:
$$
\Delta + \sum_{y} \epsilon_{y} \equiv 0 \,\,\ {\rm mod}\,\,\ 2.
\eqn\lastes
$$

At this point we have all the information that we  need to compute the
polynomial invariants for $SU(2)$ monopoles on K\"ahler, spin
four-manifolds. Notice that the result will involve unknown constants which
should be fixed by comparing to mathematical computations of these
invariants. These  constants are universal, in the sense that they depend
only on  the dynamics of the physical theory (as shown in [\wjmp]) and not
on the particular manifold we are considering. If we denote $C=C_1$, $\gamma
= \gamma_1$, $\langle 1  |{\cal O}|1 \rangle = \langle {\cal O} \rangle$ and
$\langle 1 +|V|1+ \rangle = \langle   V\rangle$, the expression for the
polynomial invariants reads:
$$
\eqalign { &C \Bigg( {\rm exp} (\gamma v^2
+\mu \langle {\cal O} \rangle)
\prod_{y}\Big( {\rm exp}(
\langle V \rangle \phi_{y}) + t_{y}{\rm exp} (-\langle V
\rangle \phi_{y}) \Big) \cr &+(-1)^{\Delta}{\rm e}^{-{\pi i \over 12}\sigma}
{\rm exp}\Big( {\rm e}^{-{2\pi i
\over 3}}(\gamma v^2 +\mu \langle{\cal O} \rangle) \Big)
\prod_{y}\Big( {\rm exp}( {\rm
e}^{-{\pi i \over 3}} \langle V \rangle \phi_{y}) + t_{y}{\rm exp}( -{\rm
e}^{-{\pi i \over 3}}\langle V
\rangle\phi_{y} \Big) \cr
&+{\rm e}^{-{\pi i \over 6}\sigma}
{\rm exp}\Big( {\rm e}^{-{4\pi i
\over 3}}(\gamma v^2 +\mu \langle{\cal O} \rangle) \Big)
\prod_{y}\Big( {\rm exp}( {\rm
e}^{-{2\pi i \over 3}} \langle V \rangle \phi_{y}) + t_{y}{\rm exp}( -{\rm
e}^{-{2\pi i \over 3}}\langle V
\rangle\phi_{y} \Big) \Bigg). \cr}
\eqn\polyone
$$

In order to check some of the properties of \polyone\ we will express
it in a more convenient way.
Notice that because of \late\ and \lastes\ we can
extract the factor $t_{y}$ in the second
summand of \polyone\ and cancel the factor $(-1)^{\Delta}$. Using some
straightforward algebra and the fact that  $\sigma \equiv 0 \,\
{\rm mod}\,\ 8$, \polyone\ can be rewritten as:
$$
\eqalign{
&\langle {\rm exp} (\sum_{a}\alpha_{a}I(\Sigma_{a})+\mu {\cal O})
\rangle\cr
&=C \Bigg( {\rm exp}
(\gamma v^2 +\mu \langle {\cal O} \rangle) \prod_{y}\Big( {\rm
exp}( \langle V \rangle \phi_{y}) + t_{y}{\rm
exp} (-\langle V \rangle \phi_{y}) \Big) \cr
&+{\rm e}^{-{\pi i \over 6}\sigma}
{\rm exp}\Big( -{\rm e}^{-{\pi i
\over 3}}(\gamma v^2 +\mu \langle{\cal O} \rangle)
\Big) \prod_{y}\Big( {\rm exp}( {\rm
e}^{-{2\pi i \over 3}} \langle V \rangle \phi_{y}) + t_{y}{\rm exp}( -{\rm
e}^{-{2\pi i \over 3}}\langle V
\rangle\phi_{y} \Big) \cr
&+{\rm e}^{-{\pi i \over 3}\sigma}
{\rm exp}\Big( -{\rm e}^{{\pi i
\over 3}}(\gamma v^2 +\mu \langle{\cal O} \rangle)
\Big) \prod_{y}\Big( {\rm exp}( {\rm
e}^{-{4\pi i \over 3}} \langle V \rangle \phi_{y}) + t_{y}{\rm exp}( -{\rm
e}^{-{4\pi i \over 3}}\langle V
\rangle\phi_{y} \Big) \Bigg) \cr}
\eqn\fer
$$
where the second summand of
\polyone\ is now the last one. This is our final expression for the
polynomial invariants associated to $SU(2)$ monopoles on K\"ahler, spin
manifolds whose canonical divisor can be written as in \canonico. The result
is obviously real,  as the first summand in \fer\ is real and the second one
is the complex conjugate of the third one.

Another check of \fer\ is the following. As we
noticed in sect. 2, a product of $r$
observables ${\cal O}$ and $s$ observables
$I(\Sigma)$ has ghost number $4r+2s$,
and this must equal the dimension of the moduli
space for some instanton number $k$. In terms
of $\Delta$ we have the selection rule \selrule:
$$
4r+2s={\rm dim} \,\ { \cal M }_{\rm NA}=6(k-\Delta )-{\sigma \over 2},
\eqn\dimension
$$
{\it i.e.}, if we suppose that the
$\alpha_a$ are of degree $2$ and $\mu$ of degree
$4$, in the expansion of \fer\ we
can only find terms whose degree is congruent to
$-\sigma /2$ mod $6$.
This is easily checked. If we consider the terms of fixed degree
$4r+2s$ we see that they can be grouped
in terms with the same coefficient, given by:
$$
1+{\rm e}^{-{\pi i \over 6}\sigma}{\rm e}^{-{\pi i \over 3}
(4r+2s)}+{\rm e}^{-{\pi i \over
3}\sigma}{\rm e}^{-2{\pi i \over 3}(4r+2s)}.
\eqn\bilbao
$$
This is a geometrical series whose sum
is zero unless ${\rm e}^{-{\pi i \over 6}\sigma-{\pi i
\over 3}(4r+2s)}=1$, which gives
precisely the condition we were looking for. Notice
that to obtain the well-behaved
expression \fer\ from \polyone\ the key point is that
the contributions from the
cosmic string theory have the form \cosmicomicas. This is
what allows to drop out the
factor $(-1)^{\Delta}$ which comes from the bulk structure
and suggests that the pattern
of bifurcation of vacua along the cosmic string is the
right one.

Another point of interest is that,
according to our expression \fer,
the generating function for the correlation functions $f=\langle {\rm exp}
(\sum_{a}\alpha_{a}I(\Sigma_{a})+\mu {\cal O}) \rangle $
verifies the equation:
$$
{\partial^3 f \over \partial \mu^3}={\langle {\cal O} \rangle }^3 f,
\eqn\tiposimple$$
which seems to be the adequate
generalization to our moduli problem of the simple type
condition which appears in Donaldson theory
\REF\km{P.B. Kronheimer and T.S. Mrowka \journal\bams&30(94)215}
[\km]. Physically, the order of
this equation is clearly related to the number of
singularities which appear
in the quantum moduli space of vacua. It would be
interesting to have a clear picture of the mathematical meaning of this
generalized simple type condition
as well as to know what is the form it takes in
other moduli problems.

\section{General spin manifolds}

In the previous section we have
computed the polynomial invariants for $SU(2)$
monopoles on K\"ahler, spin manifolds.
The fact that we have a K\"ahler structure
allows one, as we have seen in sect. 2, to perform the computation in the
$N=1$ theory. In this section we will show that one can
use electric-magnetic duality and the
$U(1)_{\cal R}$ symmetry of the
original $N=2$ theory to obtain expressions which are
valid on a general spin manifold $X$, as it happens in Donaldson theory
[\mfm]. As it has been shown in
[\wspure, \wsmatter], at every point in the quantum
moduli space of vacua of the $N=2$ Yang-Mills theory
(the $u$-plane) there is a
low-energy abelian $N=2$ effective
theory which can also be twisted to give a topological
field theory. At a generic point the only light degree of freedom is the
$U(1)$ gauge field which survives after gauge symmetry breaking, and the
twisting of this theory would give
(as it should be clear from sect. 2) the moduli
problem of abelian instantons on $X$.
At the singularities new massless states
(monopoles or dyons) appear
which must be included in the low-energy lagrangian. The
resulting effective theory is $N=2$ QED with a certain number of massless
hypermultiplets. For the pure $N=2$ Yang-Mills theory and the theory with
$N_f=1$ which we have been considering, there is a single hypermultiplet
at every singularity [\wsmatter]. In these cases,
the twisted theory near these points gives the
abelian monopole equations of [\mfm]. Such a theory
 has been constructed in
[\toplag]. In principle, when computing a correlation
function of the original,
``microscopic" twisted theory, one should
integrate over the $u$-plane. However, the moduli
problem in Donaldson theory and in the non-abelian monopole theory (as it
has been argued in [\nonab])
are well defined only for manifolds with $b_2^{+}>1$. This
condition means that there are
no abelian instantons on $X$ for a generic metric, and
therefore one expect contributions only from the singularities, as
the moduli space of the twisted effective, ``macroscopic" theory is
empty for the other points in the $u$-plane. This is consistent with
the $N=1$ point of view. Once we know the contribution from one of
the singularities, the other contributions can be obtained through
the underlying  microscopic $U(1)_{\cal R}$ symmetry of the $N=2$ theory.

Let us implement this picture in our problem.
The quantum moduli space of vacua has
three singularities, as we have
recalled in sect. 2. Each of them corresponds to a
single state becoming massless.
The charges of the three different states are $(n_{m},
n_{e})=(1,0)$, $(1,1)$, $(1,2)$, where $n_m$ denotes the magnetic charge
and $n_e$ the electric one [\wsmatter].
Consider the singularity associated to the
magnetic monopole, with charge $(1,0)$.
The low-energy effective theory after twisting
gives the moduli problem of abelian monopoles,
and as the observables of the theory
are those of the pure Yang-Mills case,
we should expect that the contribution from this
singularity to a correlation
function is the same which appears in the $N=2$ theory
without matter coming from the singularity
at $u=\Lambda_{0}^2$. Recall from [\mfm]
that the abelian monopole theory is
defined in terms of a complex line bundle $L$ (in the
spin case) or equivalently by a class $x=-2c_1(L)$. When the moduli space
associated to the abelian monopole
equation has zero dimension, $x$ satisfies:
$$
x^{2}=2\chi + 3\sigma,
\eqn\basicas$$
and the partition function of this
theory is denoted by $n_x$. The $x$ such that $n_x
\not= 0$ are called basic
classes, and they must verify the condition \basicas.
According to [\mfm], the contribution
from the vacua at $u=\Lambda_{0}^2$ is given by:
$$
{\rm exp} (\gamma v^2 +\mu \langle
{\cal O} \rangle) \sum_{x} n_x {\rm e}^{\langle V
\rangle v \cdot x},
\eqn\unvacio
$$
where $v \cdot x=\sum_{a}
\alpha_{a} \sharp(\Sigma_{a}\cap x)$ and we have included the
new universal constants which
also appear in \fer. The sum is over all the basic
classes. Now, to get the contributions
from the other two vacua we can use the $U(1)_{\cal R}$
symmetry given in \lola\ and \maslola. The resulting ${\bf Z}_6$
anomaly-free discrete
subgroup in \enedos.
 Notice that if we use \enedos\
the transformation of the order parameter $u$ is $u \rightarrow {\rm
e}^{-{4 \pi i \over 3}}u$.
This is still a ${\bf Z}_3$ symmetry of the $u$-plane
which goes through all the singularities, and we won't need to implement the
additional symmetry \parida.

At this point the computation becomes very similar to the one we did for the
$N=1$ theory. First we must take into account the gravitational contribution
of the anomaly and the geometrical character of the fields after twisting.
The fields $\lambda^{1}$, $\lambda^{2}$,  ${\overline \lambda_{1}}$ and
${\overline \lambda_{2}}$ give now the whole ASD complex \asd\ and the
anomaly is the square of \fase. For the matter fermions the anomaly is given
by ${2 \pi i \over 3} {\rm index}\,\ D $. The total contribution is:
$$
{\rm e}^{ {\pi i \over 3}\{ 8k-{3 \over 2}(\chi
+\sigma)\}- {\pi i \over 3}(2k+{\sigma \over 2})}={\rm e}^{-{\pi i \sigma
\over 6}}.
\eqn\vitoria
$$
This is the anomaly we obtained for the third $N=1$ vacuum, for it is the
square of \anomaly.
We see that, as it happens in the pure $N=2$ Yang-Mills theory
[\wjmp], the $U(1)_{\cal R}$
symmetries of the $N=1$ and the $N=2$ theory, which are
certainly different,
work in such a way that after twisting one obtains the same
contribution for the gravitational
part of the anomaly.

To implement the symmetry under consideration in
the observables we need the
action of the generator of \enedos, call it again $\alpha$,
on them. One obtains:
$$
\alpha {\cal O} \alpha^{-1} =
-{\rm e}^{\pi i \over 3} {\cal O},\,\,\,\,\ \alpha
I(\Sigma) \alpha^{-1} = {\rm e}^{2\pi i \over 3} I(\Sigma) .
\eqn\obenedos
$$
Now we can apply these transformations to
\unvacio, as we did in the K\"ahler case, to
obtain:
$$\eqalign {
&\langle {\rm exp} (\sum_{a}\alpha_{a}I(\Sigma_{a})+\mu {\cal O})
\rangle\cr
&=C \Bigg( {\rm exp} (\gamma v^2
+\mu \langle {\cal O} \rangle) \sum_{x} n_x {\rm
exp}(\langle V \rangle v \cdot x)  \cr
&+{\rm e}^{-{\pi i \over 6}\sigma}
{\rm exp}\Big( -{\rm e}^{-{\pi i \over 3}}
(\gamma v^2 +\mu \langle{\cal O} \rangle)
\Big) \sum_{x} n_x {\rm exp}({\rm e}^{-{2\pi i \over 3}} \langle V
\rangle v \cdot x) \cr
&+{\rm e}^{-{\pi i \over 3}\sigma}
{\rm exp}\Big( -{\rm e}^{{\pi i \over 3}}
(\gamma v^2 +\mu \langle{\cal O} \rangle)
\Big) \sum_{x} n_x {\rm exp}({\rm e}^{-{4\pi i \over 3}} \langle V
\rangle v \cdot x) \Bigg). \cr}
\eqn\fernanda
$$
This is our final expression for
the polynomial invariants associated to $SU(2)$
monopoles on a spin manifold $X$ with $b_{2}^{+}>1$.

Using the abelian monopole
equations it is easy to show that,
when $X$ is K\"ahler and its canonical divisor is of
the form \canonico\ with disjoint $C_y$, one
recovers \fer\ from \fernanda. In such a situation
the basic classes are given by [\mfm]:
 $$
x_{(\rho_1, \cdots, \rho_n)}= \sum_{y} \rho_{y}[C_y],
\eqn\burgos
$$
and each $\rho_{y}=\pm 1$. The corresponding $n_x$ are:
$$
n_x=\prod_{y} t_{y}^{s_y} ,
\eqn\kahlerclases
$$
where $s_y=(1-\rho_{y})/2$ and
the $t_y$ are given in \late. A simple computation leads to,
$$
\sum_{x} n_x {\rm
exp}(\langle V \rangle v \cdot x)=\sum_{\rho_y}\prod_{y} t_{y}^{s_y}{\rm
exp}(\sum_{y}\langle V \rangle \rho_y \phi_{y})=\prod_{y}\Big( {\rm
exp}( \langle V \rangle \phi_{y}) + t_{y}{\rm
exp} (-\langle V \rangle \phi_{y}) \Big),
\eqn\alava
$$
which shows that in this case
\fernanda\ becomes \fer.

Our last comment concerns the
r\^ole of \tiposimple\ in the
above computation. As it is argued in [\mfm] for
Donaldson theory, the simple type
condition guarantees that the operator ${\rm
Tr}(B^{\dagger})^2$ can be replaced by $c$-numbers in the evaluation of the
correlation functions, and there are no contributions coming from other
operators of the effective theory.
The equivalent condition in our case is certainly
\tiposimple, and therefore
the expression \fernanda\ should be valid for spin manifolds
with $b_{2}^{+}>1$  whose polynomial invariants verify this constraint.

 \endpage
\chapter{The massive theory}

In this section
we will make some observations
concerning $N=2$ QCD with one massive hypermultiplet. The
superpotential \superab\ now becomes:
$$
W={\sqrt 2}{\widetilde Q}\Phi Q + m{\widetilde Q} Q.
\eqn\supmass
$$
This theory can be twisted as in section 2.3,
but the resulting theory is not
topological because of the mass term, and can be understood in fact as a
deformation of the  theory we have been studying. Notice that the
$U(1)_{\cal R}$
 symmetry of the non-massive  theory is completely broken by the presence of
the mass term, and therefore  one would expect that the quantum moduli space
of this theory has a singularity  structure very different from the original
one. These singularities can be analyzed along the lines proposed in
[\wsmatter].

Classically the $B$ field
gets an expectation value
characterized by a complex parameter $a$, and one can
write $B=a\sigma_3$ with
$\sigma_3={\rm diag}(1,-1)$. If we write $Q=(Q^1, Q^2)$,
${\widetilde Q}=({\widetilde Q}^1,{\widetilde Q}^2)$
and expand around this vacuum, it
is easy to see from
\supmass\ that the first component of the hypermultiplet gets a mass
$m+{\sqrt 2}a$, while
the second one gets a mass $m-{\sqrt 2}a$. One finds a
classical singularity at $a=-m/{\sqrt 2}$ where the first component of the
hypermultiplet become massless: as we
have a new light degree of freedom, the
description based
on the pure $N=2$ abelian theory breaks down. If we consider that $m
\gg \Lambda_1$, where $\Lambda_1$ is the dynamically generated
mass of the theory, the
singularity is in the
semiclassical region (because $u \approx 2a^2=m^2 \gg \Lambda_1$)
and it persists in the quantum theory.
For $u \ll m^2$ all the quarks are massive and
can be integrated out. The
low-energy theory is the pure $N=2$ Yang-Mills, which has
two singularities at $u=\pm \Lambda_0 ^2$
[\wspure], where $\Lambda_0$ is the scale of
the low-energy theory and is
related to $\Lambda_1$ by $\Lambda_0^4=m\Lambda_1^3$
[\wsmatter]. The conclusion of this
analysis is that the massive theory has three
singularities, as the non-massive one,
and obviously there is not a discrete symmetry
coming from an $U(1)_{\cal R}$ relating them.

If we perturb the massive  theory with a mass term for the $\Phi$
superfield, as we have done in the computation of the polynomial invariants
for the non-massive theory, we obtain an $N=1$ theory with three vacua
coming from the singularities in the quantum moduli  space of the underlying
$N=2$ theory. Here we will obtain these three vacua for the twisted  massive
theory on a K\"ahler, spin manifold  $X$. Instead of considering the
effective theory,  as in sect. 3, we will follow the procedure used in [\wv]
to obtain similar  issues in $N=4$
Yang-Mills theory, and we will analyze the
$N=1$ superpotential  which is obtained after perturbation. Notice that, as
the mass term for the matter hypermultiplet  is $Q'_1$-closed (following the
same kind of arguments used in sect. 2), we have the same situation with
respect to the topological symmetry that the one we  had in the non-massive
theory. The only term breaking the topological invariance of  the resulting
$N=1$ theory will be again the mass term.

Recall that, as we are on a
K\"ahler manifold, the terms in the superpotential must
be $(2,0)$ forms. The mass
perturbation for the $\Phi$ superfield must be done with
a global $\omega$, which is a
holomorphic section of the canonical bundle
$ K$ over $X$. In this paper we take it satisfying
\canonico. The superpotential reads:
$$
W={\widetilde Q}\Phi Q + m {\widetilde Q} Q + \omega{\rm Tr} \Phi^2.
\eqn\superraro
$$
In order to obtain the critical points of this function
we have to solve the following equations:
$$
\eqalign{
&{\partial W \over \partial Q_a}={\sqrt
2}{\widetilde Q}_b \Phi_{ab}+m{\widetilde Q}_a=0,\cr
 &{\partial W \over \partial
{\widetilde Q_a} }={\sqrt 2}
\Phi_{ab}Q_a +m Q_a=0, \cr &{\partial W \over \partial
\Phi_{ab}}={\sqrt 2}{\widetilde Q}_a Q_b
+2{\overline \omega} \Phi_{ab}=0,\,\,\,\,\,\ a
\not=b, \cr &{\partial W
\over \partial \Phi_{11} }= {\sqrt 2} ({\widetilde Q}_1
Q_1-{\widetilde Q}_2 Q_2)+4
\omega \Phi_{11}=0.\cr}
\eqn\criticos
$$
An obvious solution of the equations \criticos\ is the trivial one, with
$\Phi=Q=\widetilde Q =0$.
This corresponds to the trivial embedding solution in [\wv],
and gives at low energies
the pure $N=1$ Yang-Mills theory, with two different vacua.
This is the same structure we found in the quantum
moduli space of vacua of the
$N=2$ massive theory for the
low-energy behaviour. Because of the third singularity,
we should expect a non-trivial
critical point for \superraro. The first equation of
\criticos\ tells us that $Q$ is
an eigenvector of $\Phi$ with eigenvalue $-{\sqrt
2}m/2$. As $\Phi$ is a traceles
matrix, their eigenvalues must be $-{\sqrt 2}m/2$
and ${\sqrt 2}m/2$. Now recall
that we must quotient the solutions of \criticos\ by
the group of complexified gauge transformations,
{\it i.e.} transformations in
$Sl(2,{\bf C})$. We can use
this gauge freedom to write $\Phi$ in diagonal form:
$$
\Phi \longrightarrow M \Phi M^{-1} = -{\sqrt 2}m/2 \left(\matrix{1&{0}\cr
                         {0}&-1\cr}\right),\,\,\,\,\,\ M \in Sl(2,{\bf C}).
\eqn\virgo$$
This means that at this
critical point $\Phi$ breaks the gauge group down to $U(1)$,
as it should be
expected from the description by means of the $N=2$ theory.

In the twisted theory on
a K\"ahler manifold we have $Q=( \alpha_1, \alpha_2) \in
K^{1/2} \otimes E$
and similarly ${\widetilde Q}=( \beta_1,  \beta_2) \in K^{1/2}
\otimes {\overline E}$
(here we denote by capital letters also the first component of
the chiral superfields $Q$, ${\widetilde Q}$ and $\Phi$).
The vacuum of the theory
must have zero action, and this requires $\Phi$ to be
holomorphic because of its kinetic energy term. But if a holomorphic
section of the adjoint gauge bundle ${\rm ad} E$
splits everywhere as in \virgo,
then $E=L\oplus L^{-1}$, with $L$ a holomorphic line
bundle. Let us analyze
the remaining equations in \criticos. After conjugation by $M$,
$Q$ and ${\widetilde Q}$ verify:
$$
\eqalign{ &{\widetilde Q}_1 Q_2 = {\widetilde Q}_2
Q_1=0,\cr &{\widetilde Q}_1 Q_1-{\widetilde Q}_2 Q_2=2m \omega. \cr}
\eqn\culos
$$
Choosing $Q_1 \not =0$,
we have ${\widetilde Q}_2=Q_2=0$, and ${\widetilde Q}_1
Q_1=2m \omega$. Because of the splitting of $E$, we have
$\alpha \equiv \alpha_1 \in K^{1/2} \otimes L$,
and $\beta \equiv \beta_1 \in K^{1/2} \otimes L^{-1}$. The
last equation in \culos\ gives:
$$
\alpha \beta =2m \omega,
\eqn\monopolos
$$
which is essentially the
(perturbed) abelian monopole equation of [\mfm] on a K\"ahler
manifold (notice also that in
order to obtain a vacuum $\alpha$ and $\beta$ must be
holomorphic, as required in [\mfm]).

The above result confirms the picture for the low-energy theories associated
to the singularities of the $N=2$ massive theory: at these points the light
degrees of freedom are a  matter hypermultiplet and the $N=2$ photon, and
the effective theory should be $N=2$ QED with one matter field. This theory,
after twisting,  gives the moduli problem of abelian monopoles, and its
vacua correspond to the solutions of these equations. This is precisely what
we have obtained as the critical points of $W$.  Notice that
 the two bulk vacua associated to the trivial solution, and corresponding to
pure $N=1$ Yang-Mills, have the same internal vacuum strucure  given by the
solution to the monopole equations.  This is because, when the canonical
divisor has the form  \canonico, the solutions of \monopolos\ correspond
precisely to the two-fold bifurcation of  each bulk vacuum along the cosmic
strings $C_y$, as it is shown in [\mfm].

One should expect
that the ``internal" vacuum
structure associated to the bulk vacua of the massive theory
are equivalent to the
ones arising in the non-massive theory. This is clear from the
$N=2$ point of view,
where the low-energy effective theories are equivalent, at least
in the limit of very large
$\Lambda$. Therefore, the analysis that we have done supports
the assumption we made in sect. 4 about
the cosmic string theory. Notice that we have
not used duality arguments in this analysis,
but rather we have checked their predictions.

\endpage
\chapter{Conclusions}

In this paper we have computed the polynomial invariants associated to the
moduli space of $SU(2)$ monopoles on four-dimensional spin manifolds, with
the monopole fields in the fundamental  representation of the gauge group.
Our computation is based on the exact results about the quantum moduli space
of vacua of the corresponding $N=2$  and $N=1$ supersymmetric theories, and
follows the lines of [\wjmp] and [\mfm]. The resulting expressions
\fernanda\ and \fer\ can be written in terms of Seiberg-Witten invariants,
and therefore the first conclusion of our analysis is that  these invariants
underlie not only Donaldson theory, but also the generalization of
 this theory presented in [\nonab]. This is a striking result, as the moduli
space of $SU(2)$  monopoles seems at first view very different from the
moduli space associated to the  abelian monopole equations. Certainly it
should be very interesting  to have a mathematical understanding of this
fact, as well as expressions for the monopole  invariants computed by
mathematical methods in order to compare them to our results.

 The picture which emerges from our computation  is that non-perturbative
methods in supersymmetric gauge theories are not only an  extremely
powerfool tool to obtain topological invariants, but also to relate very
different moduli problems in four-dimensional geometry. Notice that the
information about the quantum moduli space of vacua in  [\wspure,\wsmatter]
is obtained  integrating out the massive excitations of the original field
theory, in order to obtain low-energy effective descriptions. The
topological information of the twisted, microscopic theories seems to be
encoded in only two parameters
of the non-perturbative results: the number of
singularities in the moduli space of vacua (related by an anomaly-free
discrete subgroup) and the number of hypermultiplets becoming massless at
these singularities.  It seems that different four-dimensional moduli
problems can be in the same ``universality class"  when considered from the
point of view of the underlying supersymmetric theories.  Therefore, using
non-perturbative results in the physical theories, one should be able to
identify truly basic topological invariants characterizing a whole family of
moduli problems. According to our results, the $SU(2)$ Donaldson invariants
and the $SU(2)$ monopole invariants are both in the same class, which is
associated to the Seiberg-Witten invariants (as the topological information
that both give is encoded in the basic  classes of the manifold). In order
to explore this kind of behaviour, the first problem which should be
 addressed is the analysis of the different topological field theories which
arise from the twisted $N=2$ supersymmetric QCD. Conversely, one could check
the predictions of the physical theories by comparing them to mathematical
results. This would give a  very fruitfull arena for the interaction of
physics and mathematics which topological field theories have made possible.

\vskip1cm

\ack We would like to thank O.
Garc\'\i a-Prada, A. Klemm and S. Yankielowicz for very
helpful discussions, and E. Witten for useful correspondence.
We would also like to thank the Theory Division at CERN, where
this work was completed,
for its hospitality. This work
was supported in part by DGICYT under grant PB93-0344 and  by CICYT
under grant AEN94-0928.

\vskip1cm

{\bf Note added:} During the review of this work we have rederived the bulk
structure of the vacuum using non-perturbative $N=1$ supersymmetric methods. We
present this analysis in Appendix B. We have also included some details on the
explicit realization of the $\rho$ symmetry in Appendix A.

\endpage
\Appendix{A}
In this appendix we will make some observations concerning the parity
symmetry \parida. As this symmetry  is a consequence of the isomorphism
between the ${\bf 2}$ representation and the ${\bf {\bar 2}}$
 representation of $SU(2)$, we will construct  explicitly this isomorphism.
We will focus on the case $N_{f}=1$, although these considerations extend
inmediately to the general case.

When $N_{f}=1$, in $N=1$ language we have a chiral superfield (quark) $Q_a$
transforming in the ${\bf 2}$ of $SU(2)$ and another chiral superfield
(antiquark) $\widetilde Q_a$ transforming  in the ${\bf {\bar 2}}$. The
index $a$ is a color index. Now we can define the fields:
$$
\eqalign{
{\hat Q}^1_a=&Q_a,\cr
{\hat Q}^2_a=&(\sigma_2)_{ab}\widetilde Q_b, \cr}
\eqn\primtrans
$$
where $\sigma_2$ is a Pauli matrix. If $U \in SU(2)$, as $\sigma_2
U^{*}\sigma_2=U$, the field ${\hat Q}^2_a$ transforms also in the  ${\bf 2}$.
This is an explicit realization of the isomorphism ${\bf 2}  \simeq {\bf
{\bar 2}}$. We must also redefine the chiral superfield $\Phi$, which lives
in the adjoint  representation, in the following way:
$$
\hat {\Phi} =(\sigma_2)^{\rm T} \Phi.
$$
The new field $\hat {\Phi}$ is a symmetric matrix because ${\rm Tr} \Phi=0$.
The $N=2$ coupling
$\widetilde {Q}^{\rm T} \Phi Q$ is written in terms of the new variables as:
$$
{1 \over 2}({\hat Q}^2_a
\hat {\Phi}_{ab}{\hat Q}^1_b +{\hat Q}^1_a
\hat {\Phi}_{ab}{\hat Q}^2_b).
\eqn\acopla
$$
The $N=2$ mass term for the matter
fields involves the gauge invariant quantity $X=\widetilde {Q}_a Q_a$, which
in the new variables is written in the form,
$$
X=-i({\hat Q}^2_2 {\hat Q}^1_1 -{\hat Q}^1_2 {\hat Q}^2_1).
\eqn\masacambia
$$
The parity transformation, which
interchanges the quark and the antiquark, must be properly
understood in terms of these variables as,
$$
\rho : {\hat Q}^1 \leftrightarrow {\hat Q}^2.
\eqn\paridapendice$$
As the term \acopla\ is invariant under
\paridapendice, this is a symmetry of $N=2$ QCD with massless matter
fields. Notice however
that the $SU(2)$ singlet $X$ changes
its sign under \paridapendice, as it is obvious from \masacambia.
Therefore the $N=2$ mass term for the
quark and the antiquark changes its sign accordingly.

Another set of variables which is
useful to take into account the $\rho$ symmetry is the following:
$$
\eqalign{
Q^1=&{1 \over 2i}({\hat Q}^1-{\hat Q}^2),\cr
Q^2=&{1 \over 2}({\hat Q}^1 +{\hat Q}^2). \cr}
\eqn\variadef
$$
The $N=2$ coupling in these new variables reads as,
$$
 Q^1 \hat {\Phi} Q^1+ Q^2 \hat {\Phi} Q^,
$$
while the singlet $X$ takes the form:
$$
X=2(Q_1^1 Q_2^2-Q_2^1 Q_1^2).
$$
The parity symmetry in terms of these variables is,
$$
\eqalign{
&Q^1 \rightarrow -Q^1, \cr
&Q^2 \rightarrow Q^2. \cr}
\eqn\paridafinal
$$
Using the variables defined in \variadef\ it is easy to see that the flavour
symmetry for the $N=2$
QCD with gauge group $SU(2)$ and $N_f$ hypermultiplets is $O(2N_f)$.
\endpage

\Appendix{B}
In this appendix we will derive the vacuum structure and the pattern of
chiral symmetry breaking  of the $N=1$ theory with a mass term for the
superfield $\Phi$ using the non-perturbative methods developed in [\seiberg
-\fases]. This also shows that in certain conditions exact results for $N=1$
theories can be useful in topological computations.

The $N=1$ theory we are interested in is an $SU(2)$ theory with a quark $Q$,
an  antiquark $\widetilde{Q}$ and a triplet  $\Phi$. Apart from the minimal
gauge couplings to the Yang-Mills field, this theory has a coupling  between
these three matter fields coming from the $N=2$ supersymmetry (the last two
terms in \silvana) and also a  mass term for $\Phi$ given in \masa. The
vacua of this theory can be found as the minima of the exact superpotential,
and to obtain this we can use a technique developed in [\intin] and called
the ``integrating in" procedure. This technique  allows one to obtain the
exact superpotential for an ``upstairs" theory starting from the one of a
 ``downstairs" theory. The upstairs theory differs from the downstairs
theory in that it  contains an additional matter field. In our case we can
take as the downstairs theory the $SU(2)$  theory with a quark and an
antiquark, whose exact superpotential is known   [\susybreak, \vacua], and as
the additional field for the upstairs theory the chiral superfield in the
adjoint representation, $\Phi$.  To ``integrate in" the field $\Phi$ we must
consider the gauge-invariant  polynomials which include this field.
In our case they are simply,
$$
U={\rm Tr} \Phi ^2,\,\,\,\,\,\ Z=\sqrt 2 {\widetilde Q } \Phi,
Q
\eqn\gaugepol
$$
and we must turn on a tree-level superpotential:
$$
W_{\rm tree}=mU+\lambda Z.
\eqn\arbol$$
The scales $\Lambda_{d}$ of the downstairs theory and $\Lambda$ of the
upstairs theory with the  mass term in \arbol\ are related according to the
principle of simple thresholds [\intin]:
$$
\Lambda_{d}^5=\Lambda^3 m^2.
\eqn\escalas
$$
The full superpotential of the upstairs
theory with the additional tree-level term \arbol\
is given by the principle of linearity [\superp, \intin],
$$
W_f(X, U, Z, \Lambda^3, m, \lambda)=W_u(X, U, Z, \Lambda^3)+mU+\lambda Z,
\eqn\full
$$
where $X$ is the gauge-invariant polynomial of the downstairs theory,
$X={\widetilde Q} Q$,  and $W_u$ is the exact superpotential of the upstairs
theory we are looking for. If we integrate  out the field $\Phi$ and,
correspondingly, the fields $U$ and $Z$, we obtain a new superpotential:
$$
W_l(X, \Lambda^3, m, \lambda)=W_d(X,
\Lambda_{d}^5)+W_{I}(X,\Lambda^3, m, \lambda).
\eqn\abajo
$$
In this equation $W_d$ is the dynamically
generated superpotential of the downstairs theory and is
given by,
$$
W_d(X, \Lambda_{d}^5)= { \Lambda_{d}^5 \over X}.
\eqn\supabajo
$$
In \abajo\ $W_{I}$ is an additional term which must be determined using the
symmetries of the problem together with holomorphy principles and the
behaviour of the superpotential in various limits. The first contribution to
this piece comes after integrating out $\Phi$ from $W_{\rm tree}$. In this
case the result is [\fases],
$$
W_{\rm tree, d}=-{\lambda^2 \over 4m} X^2.
$$
The upstairs theory has two non-anomalous symmetries which can be used to
constrain the form of $W_{I}$,  following the methods of [\seiberg]. The
first one is a $U(1)$ symmetry under which $Q$, $\widetilde Q$, $\Phi$, $m$
and $\lambda$ have charges $2$, $2$, $-1$, $2$ and $-3$, respectively. The
other one is a
 $U(1)_{\cal R}$ symmetry with charges $1$, $1$, $-1$, $0$ and $-3$. The
invariance of the superpotential  under these symmetries and the holomorphy
determine the form of $W_{I}$:
$$
W_{I}={  X^2 \lambda^2 \over m}
f\Big( {\Lambda^3 m^3 \over X^3 \lambda^2} \Big),
$$
where $f(u)=\sum_{n=0}^\infty a_n u^n$
is an analytic function. Notice that the first term of
this expansion corresponds to
$W_{\rm tree, d}$. Now, in the
$m \rightarrow \infty$ limit, only $W_d$ survives,
and this implies that the coefficients $a_n$ in the expansion
of $f(u)$ must be zero for $n>0$.
Therefore $W_{I}=W_{\rm tree, d}$ and the superpotential
\abajo\ is given by:
$$
W_l(X, \Lambda^3, m, \lambda)
={ m^2 \Lambda^3  \over X}-{\lambda^2 \over 4m} X^2.
\eqn\respuesta
$$
The interesting thing now is
that $W_l$ is the Legendre transform of $W_u$, which can be obtained from
$$
W_n=W_l(X, \Lambda^3, m, \lambda)-mU-\lambda Z,
\eqn\casi$$
by integrating out $m$ and $\lambda$,
{\it i.e.}, by an inverse Legendre transform. The expectation
 values of these parameters are:
$$
m={ X U \over 2 \Lambda^3}
\Big( 1-{Z^2 \over X^2 U} \Big),\,\,\,\,\,\,\ \lambda=
-{ Z U \over X \Lambda^3}\Big( 1-{Z^2 \over X^2 U} \Big),
\eqn\param
$$
and substituting these values in \casi\ one
gets the superpotential of the upstairs theory [\fases]:
$$
W_u=-{ X U^2 \over 4 \Lambda^3}\Big( 1-{Z^2 \over X^2 U} \Big)^2.
\eqn\suparriba
$$
Now we want to obtain the vacua of the
$N=2$ theory perturbed by the $N=1$ mass term for $\Phi$. Because
of the principle of linearity, the
superpotential of this theory is given by \suparriba\ plus \arbol\
with $\lambda=1$ due to $N=2$ supersymmetry:
$$
W=-{ X U^2 \over 4 \Lambda^3}\Big( 1-{Z^2 \over X^2 U} \Big)^2+mU+ Z.
\eqn\supdef
$$
The equation $\partial W /\partial X =0$ gives,
$$
{Z^2 \over X^2 U}=-{1 \over 3},
$$
which together with $\partial W /\partial Z =0$ leads to,
$$
U^3=-{27 \over 256}\Lambda^6.
\eqn\uy
$$
This theory has therefore three vacua,
corresponding to the three roots of this equation. This is
in agreement with the results obtained
from the $N=2$ point of view. Finally, we have the \vev\ for the
 field $X$ in these vacua given by the roots of
$$
X^3={1 \over 2}m^3 \Lambda^3.
\eqn\quarks
$$
This non-zero \vev\ corresponds to the spontaneous breaking of the chiral
symmetry in \uonedef, as it happens  in [\susybreak]. The subgroup of ${\bf
Z}_{12}$ which preserves the \vev\ for the guge invariant order parameter is
precisely \martin. In this way we have rederived all the results about the
bulk structure  of the vacua using non-perturbative methods for $N=1$
theories.

\endpage

\refout
\end